\RequirePackage{lineno}
\documentclass[12pt,onecolumn]{article}	% originally amsart, the default for mathematical articles.
\usepackage[top=1in, bottom=1in, left=1in, right=1in]{geometry}

\usepackage{graphicx}
\usepackage{amssymb}
\usepackage{amsmath}  				
\usepackage{graphicx,float,wrapfig} 	% for embedding graphics
\usepackage{color}					% Allows picking of custom RGB colors
\usepackage[colorlinks,linktocpage=true]{hyperref}		% for hotlinking all \cite{} and \ref{}
\usepackage[font=small,labelfont=bf,textfont=sf,
labelsep=quad]{caption}				% points hyperref to top of figure, not caption

\usepackage{gensymb}
\usepackage{caption}
\usepackage{subfig}
\usepackage{authblk}
\usepackage{placeins}
\usepackage{booktabs}

\usepackage{xr-hyper}
\makeatletter
\newcommand*{\addFileDependency}[1]{% argument=file name and extension
  \typeout{(#1)}
  \@addtofilelist{#1}
  \IfFileExists{#1}{}{\typeout{No file #1.}}
}
\makeatother

\hypersetup{						% Hyperlinked reference setup; can use RGB color values
  citecolor= black,
  linkcolor=blue,
}

\usepackage{setspace}   

\usepackage [english]{babel}
\usepackage [english = american]{csquotes}
\MakeOuterQuote{"}

\doublespace                   				%   change line spacing

\usepackage[protrusion=false,expansion=false]{microtype}

\providecommand{\keywords}[1]{\textbf{Keywords: } #1}

\providecommand{\keywords}[1]
{
  \small	
  \textbf{\textit{Keywords---}} #1
}

\makeatletter
\renewcommand{\fnum@figure}{\thefigure}
\renewcommand{\thefigure}{Fig. \arabic{figure}}

\makeatother

\begin{document}

\title{Inkwell: Design and Validation of a Low-Cost Open Electricity-Free 3D Printed Device for Automated Thin Smearing of Whole Blood}

\author[1,$\dagger$]{Jerome Nowak}
\author[2,$\dagger$]{Anesta Kothari}
\author[3]{Hongquan Li}
\author[4]{Jaspreet Pannu}
\author[5]{Dani Algazi}
\author[2,6,7,8,*]{Manu Prakash}

\affil[1]{Department of Mechanical Engineering}
\affil[2]{Department of Bioengineering}
\affil[3]{Department of Electrical Engineering}
\affil[4]{Department of Medicine}
\affil[5]{Department of Computer Science}
\affil[6]{Department of Biology(courtesy)}
\affil[7]{Woods Institute of the Environment}
\affil[8]{Center for Innovation in Global Health}
\affil[]{Stanford University, Stanford, California, USA}
\affil[$\dagger$]{co-first author}
\affil[*]{To whom correspondence should be addressed: manup@stanford.edu}

\date{\today}
\maketitle
\newpage

\vspace{-7pt}

\begin{quote}
\begin{center}
\textsc{Abstract} \\
\end{center}
\footnotesize

Microscopy plays a crucial role in hematology and diagnosis of infectious diseases worldwide. For malaria alone, more than 200 million slides are read by manual microscopists every year. High quality thin blood smears are essential for subsequent microscopy examinations including malaria microscopy, but are hard to make in field settings. Current manual smearing methods lack consistency and often do not provide a uniformly dense mono-layer of red blood cells, even when prepared by trained experts. Existing devices for assisting in making thin smears are available but are limited by cost or complexity for wider use. Here we present Inkwell, a portable mechanical device capable of making high quality thin blood smears in field settings. Inkwell is simple, low-cost, does not use electricity, and requires minimal training prior to use. By utilizing passive dissipative dynamics of a spiral spring coupled to an air dashpot with a tunable valve - we demonstrate a highly tunable mechanism for constant velocity smears at prescribed angle. Inkwell is capable of producing high quality blood smears of tunable cell density with more than 12 million individually distinguishable red blood cells on a single slide. The current design, which exploits precision manufacturing of a 17 cents plastic syringe and a spring, can be printed on a standard 3D printer with overall unit cost of less than a few dollars in large quantities. We further present usability tests to confirm performance over 10,000 unit cycle operations with no degradation in quality of the smear and demonstrate ease of use with minimal training. Inkwell enhances the broader toolbox of open innovations in diagnostics for providing high quality medical care in low and medium resource settings. Combined with rise of 3D printing, Inkwell presents an alternative to traditional centralized manufacturing and opens up distributed manufacturing of medical diagnostics in global context.

%We present preliminary measurements to assess the consistency and reliability of our low-cost prototypes as compared to the theoretical predictions from Landau-Levich thin film coating. 
% For example, expert malaria microscopy can reach < 5 parasites/µl with examination time of >30 min but the performance is seldom achieved in regular settings. 

\end{quote}   
%\end{abstract}

\newpage
%\linenumbers\modulolinenumbers[1]
%\resetlinenumber[1]
%=========================================================================================================

%\newpage
%\linenumbers\modulolinenumbers[1]
%\resetlinenumber[1]

\section{Introduction}
\label{sec:intro}
A single drop of finger prick blood from a patient contains over 10 million blood cells. Traditionally, this drop of blood (roughly 2-4 $\mu$l in volume) is spread across a glass slide as a thin film, commonly referred to as a "thin blood smear". As a standard practice in hematology, thin smears are used across clinical diagnostics including the current gold standard test for malaria diagnosis. Across primary, secondary and tertiary hospitals around the world, a patient finger prick blood based thin smear is fixed, stained and placed under a microscope to identify a number of parasites including malaria, \cite{world2016malaria} but also babesia, loa loa, microfilariae and trypanosomes. Detecting malaria parasites early and identifying species with accuracy and sensitivity is key to controlling the disease. The ability to do this depends highly on the quality of the smears and staining \cite{sori2018external, mutabazi2021assessment} and the skill of the microscopist \cite{guerin2002malaria}. In particular, in many cases it is essential to have clear views of a single layer of red blood cells (RBCs), called the monolayer. Indeed this monolayer is useful firstly for identifying species to monitor the disease \cite{world2010basic_tutor} and secondly to diagnose late stages of infections when the parasitemia is high. Furthermore, there is a rise of automated microscopes and computer-vision based parasitemia measurement \cite{li2019octopi, das2022field}, which require consistent thin smear preparation and large monolayers of individual red blood cells for effective model training and good detection performance.

\begin{figure}[!h]%
\centering
\includegraphics[width=500pt]{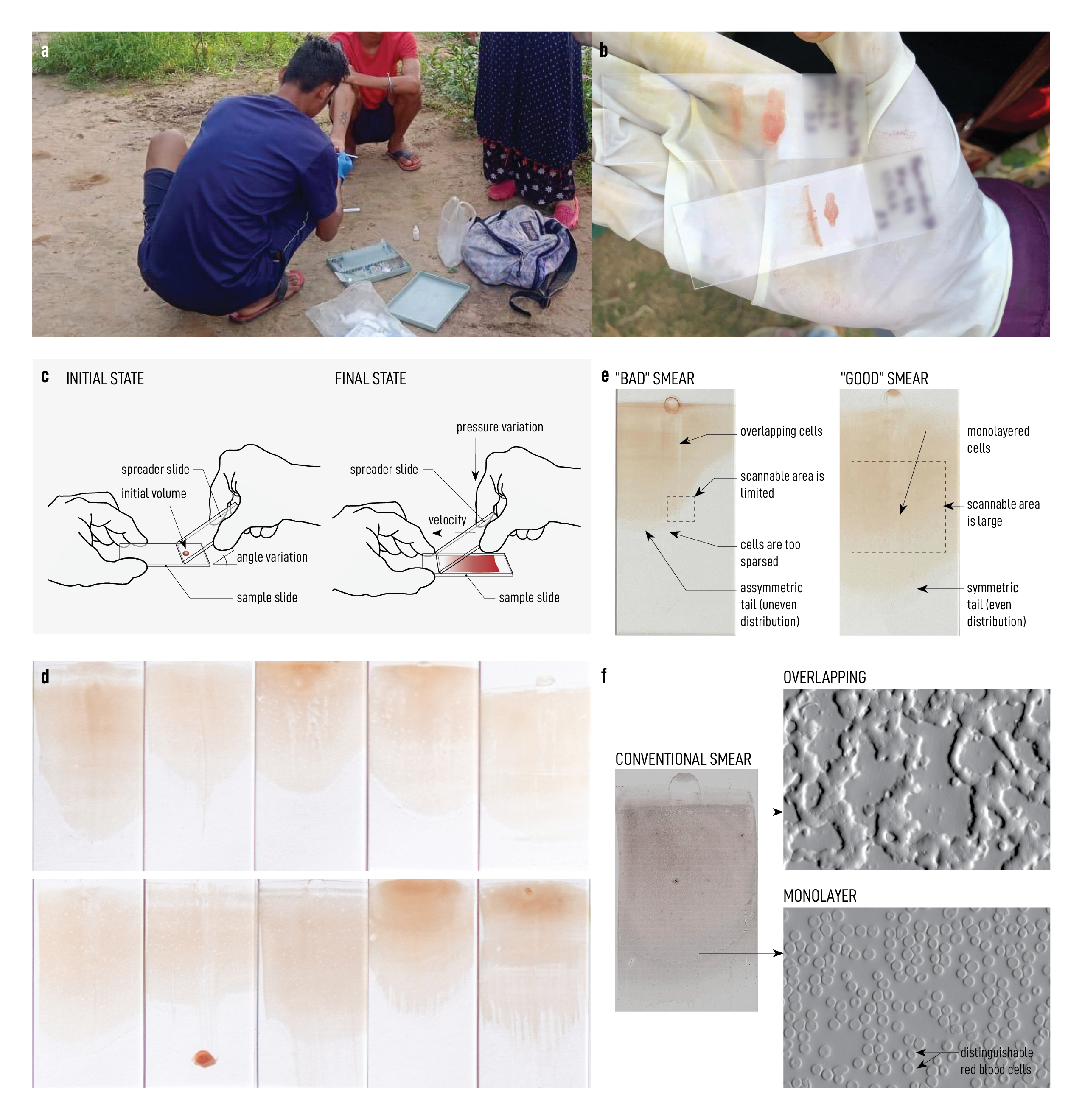}
\caption{\textbf{Variability commonly found in the quality of thin smears made by hand across different users.} [a] Field site conditions where healthcare workers visit patients door to door to collect blood samples and produce thin blood smears at the point of care. [b] Examples of typical smears produced in the field - many of which are of variable quality, resulting in inaccurate diagnoses for malaria or unreadable slides. [c] Schematic diagram showing how thin smears are made conventionally (by hand): first deposit a droplet of blood onto a sample slide, then make contact between the droplet and a spreader slide, and finally push the spreader slide across the sample slide to produce the smear. This manual method often yields in variability in angle between the two slides, velocity of the spreader slide, and pressure between the two slides, along with differences in initial droplet volume. [d] Thin smears produced by hand by lab members with varying levels of smear-making experience. [e] Some common characteristics that distinguish "good" from "bad" smears include symmetry and evenness of the smear (yielding to even distribution of monolayered cells). [f] In conventional smears, there's typically a zone with some overlapping cells and another with monolayered cells [f]. "Good" smears typically have larger monolayer zones, whereas "bad" smears are dominated with zones of overlapping cells.}
\label{fig1}
\end{figure}

However, smears produced in labs and in the field vary widely in quality due to lack of training and poor conditions for making the smears \cite{sori2018external, maguire2006production, kahama2011low, harchut2013over, ngasala2019evaluation}, which constitutes a hindrance to manual diagnosis and a barrier to automated diagnosis. Field collection of blood smears also plays an integral role in quality management, where smears are collected and centrally stored for later evaluation. Low quality smears significantly affect the functioning of any of these quality control programs. Often in remote field sites, healthcare workers need to collect and smear blood samples outside on the ground, without a stable horizontal surface on which to work [\ref{fig1}a, b]. The variability is seen in the thickness, length, symmetry, uniformity, and position of the blood film, among other aspects [\ref{fig1}b, d]. Even after a long training period, most health care workers find it difficult to make perfect thin-smears, specially in primary health care and field settings [\ref{fig1}a, b].  

Conventional smearing methods typically involve 2 glass slides: a sample slide and a spreader slide. A drop of blood is deposited onto the sample slide, then the spreader slide is brought in contact with the droplet at an angle typically between 30º and 45º such that a meniscus spanning the whole width of the slide is formed between the two glass slides. While the sample slide remains stationary, the spreader slide is pushed away from the meniscus tail to form a thin blood smear on the sample slide  \cite{CDC_thinsmear, world2010basic_learner, world2016collection} [\ref{fig1}c]. The various parameters linked to how the technician holds and moves the spreader slide, including angle of inclination, velocity and direction of motion, and whether full contact between the two slides remains constant lead to the high variability seen in smears made by hand and in the field. If the spreader slide is pushed in the other direction, towards the blood meniscus; blood cells can be sheared leading to unwanted cell lysis. 

Within a single conventional smear made by hand, the thickness of deposited blood tapers down from a thick film of overlapping cells at the start to a monolayer of cells at the finish [\ref{fig1}f]. This ultimate "feathered edge," a rounded, lighter colored area at the end of the smear, is often desired. Such an edge indicates symmetric pressure and motion and good wetting until the entire droplet was spread \cite{adewoyin2014peripheral}, resulting in a monolayer of cells ideal for parasitemia identification. What constitutes as a "good" smear is having a large zone of monolayered cells, while a "bad" smear contains largely of overlapping of cells (which makes parasite identification impossible) or an overly-sparse distribution of cells (which may not give a good representative of the sample) [\ref{fig1}e]. Although additional training can improve a technician's ability to make good smears \cite{moura2014impact}, having an accessible method or device that can consistently produce good smears regardless of technical skills can significantly improve the outcome of field diagnostics. 

Currently, there are a few products that are commercially available which addresses this need to produce consistently good smears. The Perfect Smear tool \cite{perfectsmear} is a flexible disposable plastic tool used as a spreader (in place of a glass slide) that can potentially produce good-quality smears with each use. However, it still relies on the technician's experienced dexterous hands to produce a good smear. Other devices which don't rely on technician dexterity exist on the market, like Autoslide and Hemaprep \cite{saad2020cross}, however their price of hundreds to thousands of USD is prohibitive for many markets. Aside from commercial products, Autoheam is a pair of open-source devices (one manual version and the other motorized) that were recently developed to address a similar problem. However, the manual version is still not suitable for standardizing smear-making because it still relies on the user to control the velocity of the spreader slide by hand, while the motorized version requires electricity and costs over \$50 \cite{mcdermott2022autohaem}. Furthermore, in the study of Autoheam\cite{mcdermott2022autohaem}, diluted blood instead of regular whole blood was used, limiting the direct translation of smearing parameters for clinical use. Other solutions, such as the Sysmex Hemoslider \cite{sysmex}, have been incorporated into their proprietary automated smear and staining machine (the RAL Smearbox \cite{sysmex2}), which can also represent a prohibitive investment. Therefore despite all the advancements, there is still no widely applicable, cost effective, efficient and accessible solution in this problem space.

Our work addresses the gap in affordable solutions to create high-quality smears  without extensive technical training. In particular we focus on thin smears where at least 75\% of the slide is covered in a monolayer suitable for inspection by a microscopist. Smears that contain this large coverage of monolayer are suited for emerging whole slide scanners that use computer vision for blood smear examination and quantification \cite{li2019octopi}. We first present a systematic experimental investigation of physical parameters that are optimal for creating dense monolayer smears (thin film spreading) from human whole blood. This provides a good understanding of the physical phenomena at play and helps to accurately identify the requirements for high-quality smear-making to be used in designing a cost effective and simple solutions. We then describe the complete design and evaluation of Inkwell: an open-source, low-cost, simple device that can reliably produce such smears without the need for electricity nor extensive operator training. Utilizing a 17 cent syringe and a 3D printed parts, Inkwell can be made for a few dollars in parts and capable of smearing 12 million individual cells on a single standard glass slide. We demonstrate that Inkwell can operate for 10,000 cycles continuously without any degradation in performance. 

\section*{Results}

\subsection*{Observing the blood coating process}
Macroscopically, there are three phases in the making of a thin blood smear. First, once contact between the drop of blood and the spreader slide occurs, capillary action spreads the drop across the wedge formed between the two static pieces of glass. Next, the spreader slide is pushed and the reservoir of blood travels with it, leaving behind a wet thin blood film whose thickness is dependent on slide angle and speed of actuation. Lastly, the wet blood film dries out with blood cells setting on the glass slide and it is ready for fixing.

\begin{figure*}[!h]%
\centering
\includegraphics[width=500pt]{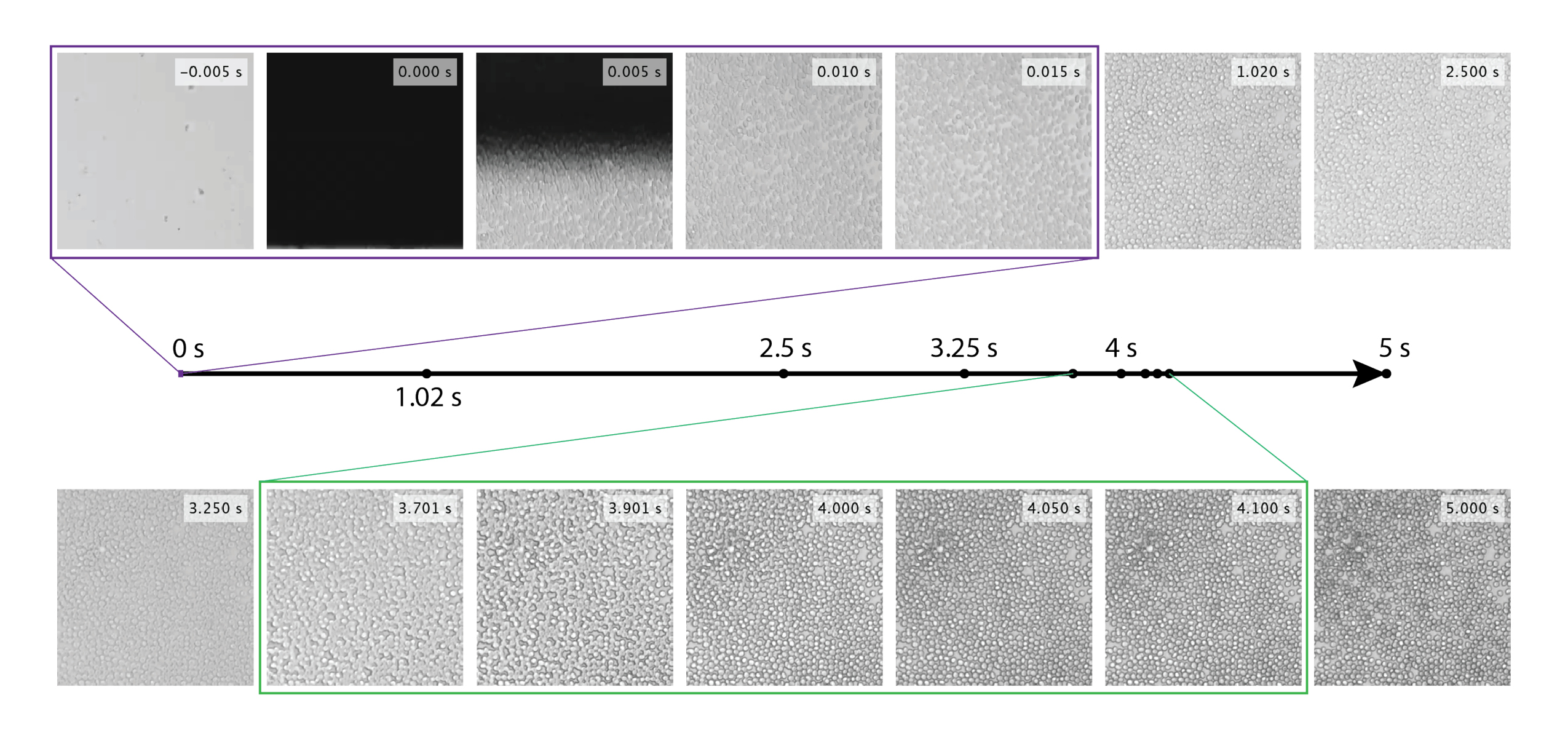}
\caption{\textbf{Blood film deposition process captured on a high-speed camera.} An array of images from the high-speed camera at various time points given an initial droplet volume of 3uL and with a velocity of 25mm/s. The images show the motion of the viscous red blood cells at their initial, settling, and dried states, clearly demonstrating various phases associated with the dynamics of thin film smear formation.}
\label{fig2}
\end{figure*}

Using a high-speed camera under a microscope (see Materials and Methods), we first imaged this process at high temporal and spatial resolution [\ref{fig2}] and supplementary video S1 A and B. It is clear from this dataset that blood cells are not immediately deposited on the slide and are floating in the fluid thin film. Two key time scales can be observed during this deposition process. Firstly, the blood reservoir traveling with the spreader slide leaves a \emph{suspension} of cells and plasma as a thin film of specific thickness behind. Secondly, with the pull of gravity and the evaporation of plasma, the blood cells quickly settle onto the slide in the next five to thirty seconds (duration depends on the layer height, blood viscosity, and ambient temperature and humidity amongst other parameters). We visualize this process using high speed imaging and also repeat this experiment for a range of spreading velocity and total blood volume. For details see supplementary videos S1 A and B and on \ref{fig2}.

\subsection*{Physical regime for thin film coatings}
The process of depositing thin films on solid substrates has been extensively studied \cite{britten1993simple} for both Newtonian and Non-Newtonian fluids. Two regimes can been identified as a function of deposition velocity $v_0$: an evaporative regime at low velocities (where the coating flow is driven mainly by solvent evaporation in the liquid meniscus), and Landau-Levich regime at higher velocities (where the coating flow is mainly driven by viscous forces) \cite{le2009convective}. From high speed imaging described above, it is clear that blood thin smear formation falls broadly in Landau-Levich\cite{reznik2009entrainment,mayer2012landau} limit with separation of time scale between evaporation (longer time scale) from deposition (shorter time scale). 

%todo Jerome: fit in papers
% Simulation of thin film coating with reflow in the meniscus, similar to blood smears, finds wegde-shaped film

% Experimental visualisation of Landau-Levitch flow with vertical plate draw
%\cite{mayer2012landau}

%Todo Jerome: acknowledge that our smears look flat, so assumptions from Russian paper may not apply to our case

Multi-scale modeling of blood hydrodynamics presents two challenges: blood is a complex non-Newtonian fluid where individual red blood cells, white blood cells and platelets are suspended in a plasma - and interact  colloid where particles (red blood cells, white blood cells, and platelets) are suspended in a solvent (plasma) \cite{thurston1972viscoelasticity,errill1969rheology}. Red blood cells (RBCs) are the most numerous particles and those of greatest interest here. Once the deposited film of blood dries, each part of the slide gets populated with the cells suspended directly above it. Assuming a homogeneous mix, this means the final surface density of RBCs on the slide after evaporation is proportionate to the thickness of the deposited fluid film. Therefore understanding what processes govern the height of the deposited thin film correlates to number of RBC per unit area. 

Colloidal coating processes have been carefully studied in the evaporative regime in \cite{doumenc2016modeling}. At typical velocities and evaporation rates, blood smearing happens in the Landau-Levich regime \cite{gutenev2003liquid}. The latter paper provides a simple formulation for blood smearing, in the form of a Newtonian fluid in Landau–Levich regime smeared at constant velocity $v_0$ and angle $\phi$ between the spreader and sample slide. In steady state (i.e. infinite supply of fluid in the meniscus), the deposited film thickness is given by $h_\infty = 1.34 R \text{Ca}^{2/3}$ where $\text{Ca} = \eta v_0 / \sigma$ is the capillary number, and $R$ the radius of the meniscus. In practice, at large timescales, the volume in the meniscus shrinks as a film is drawn from it. Assuming quasi-steady-state and no-slip condition between the deposited film and the substrate yields $ h(x) = h_{\infty 0} - k x$ where $x$ is the distance traveled since the beginning of the smear and $k$ a velocity-dependent constant given by$k(\phi, v_0) = \frac{0.898}{\tan(\frac{\pi - \phi}{2}) + \frac{\phi - \pi}{2})} (\frac{\eta v_0}{\sigma})^{4/3}$. This means for a set velocity $v_0$ and angle $\phi$, a wedge-shaped deposited film is predicted. Conservation of the volume $V_0$ yields:$
    h_{\infty 0} = \sqrt{2 k(\phi, v_0) V_0 / l}$ where $l = 25$ mm is the width of the slide. Finally, the smear height is given by: $    h(x, \phi, V_0, v_0) = \sqrt{2 k(\phi, v_0) \frac{V_0}{l}} - k(\phi, v_0) x $. This model predicts that cell density increases with velocity, initial volume, and spreader angle, and that the slope of the film does not depend on the initial volume.

%Todo Jerome: refactor, and comment on comparison with our results
%On first approximation, the spreading process happens in the Landau-Levich regime of thin film coating \cite{landau1988dragging, quere1998mouillage, gutenev2003liquid}, which has been observed in the deposition of phospholipid films \cite{le2009convective} and perovskite films \cite{deng2018surfactant} with controlled thickness. In this regime, a viscous force that is proportional to the rate of deformation determines the liquid film thickness. At higher spreading velocity, the viscous force is higher and therefore more liquid is retained, resulting in a thicker film (and, given the fixed volume of blood, a shorter film).

%This may be a useful first approximation, and the following is a summary and continuation of their results.

Given a RBC volume density of $n_V = 4–6 \times 10^6$ cells/µL \cite{cheng2004complete}, we can infer the number of cells deposited per unit surface on the sample slide: $n_S = n_V h$. Furthermore, the maximum cell density without overlaps is a densely packed hexagonal lattice, with packing density $p = \frac{\pi}{2\sqrt{3}} \approx 0.907$, and a corresponding maximum cell density:
$
    n_{S,max} = \frac{p}{\pi d^2 / 4} \approx 18,000 \text{ cells / mm}^2
$
where we estimated the diameter of a deposited RBC to be $d = 8$ µm. This means that to cover a surface $S_{tot} = 25 \times 50 \text{ mm}^2$ with a dense monolayer of RBCs (about 22 million), we need at least $V_{min} = n_{S,max}/n_V \approx 3.7–5.6 \text{ µl}$ of blood. This gives us a theoretical idea of appropriate volumes that would be useful to test in lab experiments.

\subsection*{Finding optimal smearing parameters}
%\paragraph{Metrics for optimisation}
In the case of high-throughput malaria microscopy, the optimal density is the highest number of RBCs on a given slide without overlapping cells. This optimum is theoretically achieved with cells closely packed in a hexagonal lattice of around 18,000 cells/mm$^2$ (see above). Furthermore, the greater the area covered by such a dense monolayer, the more cells can be scanned in a single smear leading to higher diagnostic sensitivity in low parasitemia samples. Our goal is to find the optimal smearing parameters to maximize both cell density and slide area covered while producing little to no overlap between RBCs. 

\begin{figure*}[!h]%
\centering
\includegraphics[width=500pt]{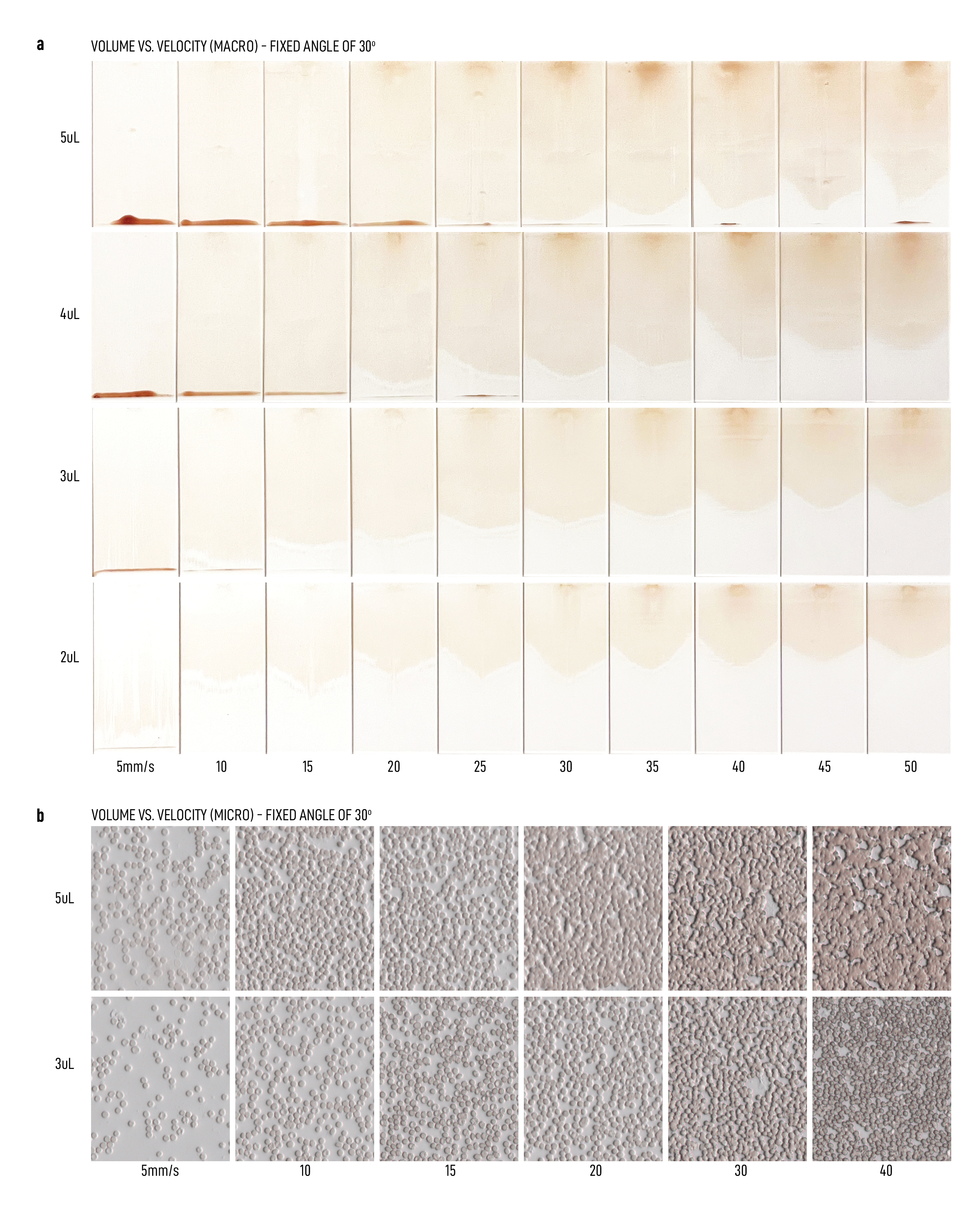}
\caption{\textbf{Analysis of the smears made under a set of controlled parameters.}
[a] A macroscopic view of the array of thin smears produced using the characterization setup, given a fixed spreader angle of 30 degrees, while varying the initial blood droplet volumes (2 µL, 3 µL, 4 µL, and 5µL) and the spreader velocities (5-50mm/s). [b] Representative microscopic fields of view for a subset of smears show the density and configuration of red blood cells given different parameters.}
\label{fig3}
\end{figure*}

To experimentally test the parameter regieme, we built a motorized smearing device and operated it at different parameters: initial blood drop volume, spreader angle, spreader velocity, and spreader contact force (Figure S4 e-g). We first  test blood droplet volumes from 2 µL to 4 µL,  spreader angles from 20º to 60º, and smearing velocities from 5 mm/s to 50 mm/s. To quantitatively evaluate the quality of the smear - we directly image the entire slide and directly map density of cells per unit area to identify optimal driving conditions. We found that increasing the blood volume, the spreader angle or the velocity results in thicker, and therefore shorter, smears. These results concur with existing practical knowledge and preliminary theory (see section above). An overview of the smears we obtained across the entire parameter space is given \ref{fig3} and figure S4b.

Although the WHO recommends holding the spreader at 30º to 45º \cite{adewoyin2014peripheral}, we found that angles below 30º, where the spreader is almost parallel to the sample slide, produced much more uniform wetting across the smear. These lower angles are less practical for an operator producing conventional manual smears, but are technically possible on our automated platform. Given a constant velocity of 15-20 mm/s and initial blood volume of 4 µL, we found 20º to work best (though we did not test lower angles for practical design reasons). About 3 to 6 µL of blood is required to cover a 50 $\times$ 25 mm$^2$ glass slide surface with a dense monolayer of RBCs up to 18,000 cells per mm$^2$ (see Materials and Methods). With 4 µl of initial blood volume and a spreader angle of 20º, we experimentally found that velocities below 15 mm/s produced sparse spacing of cells, while velocities above 20 mm/s often produced a large zone of overlapping cells. Therefore, 15 mm/s to 20 mm/s was optimal to produce near uniform smears at densities around 9,000 to 13,000 cells/mm$^2$ with almost no overlapping. While the simplified theory in previous work presented above predicts a constant negative gradient in cell density from the start to the end of the smear, we observed near uniform density across the entire smear, provided there is a sufficient initial volume of blood [\ref{fig3}a]. Thus, there is no need for non-constant velocities to compensate for the decreasing meniscus volume as blood is deposited. The quality of smears is also dependent on maintaining uniform contact between the spreader and sample slides — even a slight lift of the spreader will make the meniscus recede and hinder the wetting of the sample slide, therefore creating a non-uniform smear with large zones of overlapping cells. In tests, we determine that no additional force or pressures is needed beyond the simple gravitational force on the spreader slide on the sample slide to produce the necessary continuous contact on the resulting wedge.

%We also used theory to make predictions [supplementary], which agree with what we observe experimentally.

% Todo Jerome: refactor
%For a given blood volume, we found that increasing the velocity results in more blood deposited per distance travelled. This means that the blood film thickness tapers down more steeply from start to finish, resulting in a thicker and shorter smear. Increasing the spreader angle has a similar effect. We found that initial blood volume mostly affects how long the  doesn't affect this taper only affects how far along the smear the taper and edge will form.

\subsection*{Inkwell design: Electricity-Free Blood Smearing}

Next, we focus on the design challenge to enable optimal smearing conditions experimentally established in previous section - in an electricity free 3D printed device that is easy to use and possible to replicate anywhere. Since the device is designed to make a perfect smear every time with no user experience - we term it "inkwell". Designed for the most remote field sites, Inkwell is designed to be a low-cost and simple device that can "purely" mechanically (no batteries) recreate the exact optimal smearing parameters described above without the use of any motors or electricity. Intended to be used by non-experts, here we present design methodology for Inkwell with a focus on ease of use with minimal training for reliably and consistently producing high-quality smears of tunable thicknesses (and therefore tunable densities). As described below, all parts are made from easily accessible off-the-shelf components and simple 3D-printing approaches for reproducibility. For lack of space, we describe the design cycle and various prototypes in supplementary materials - while only focus on the final design output in detail below. 

\begin{figure*}[!h]%
\centering
\includegraphics[width=500pt]{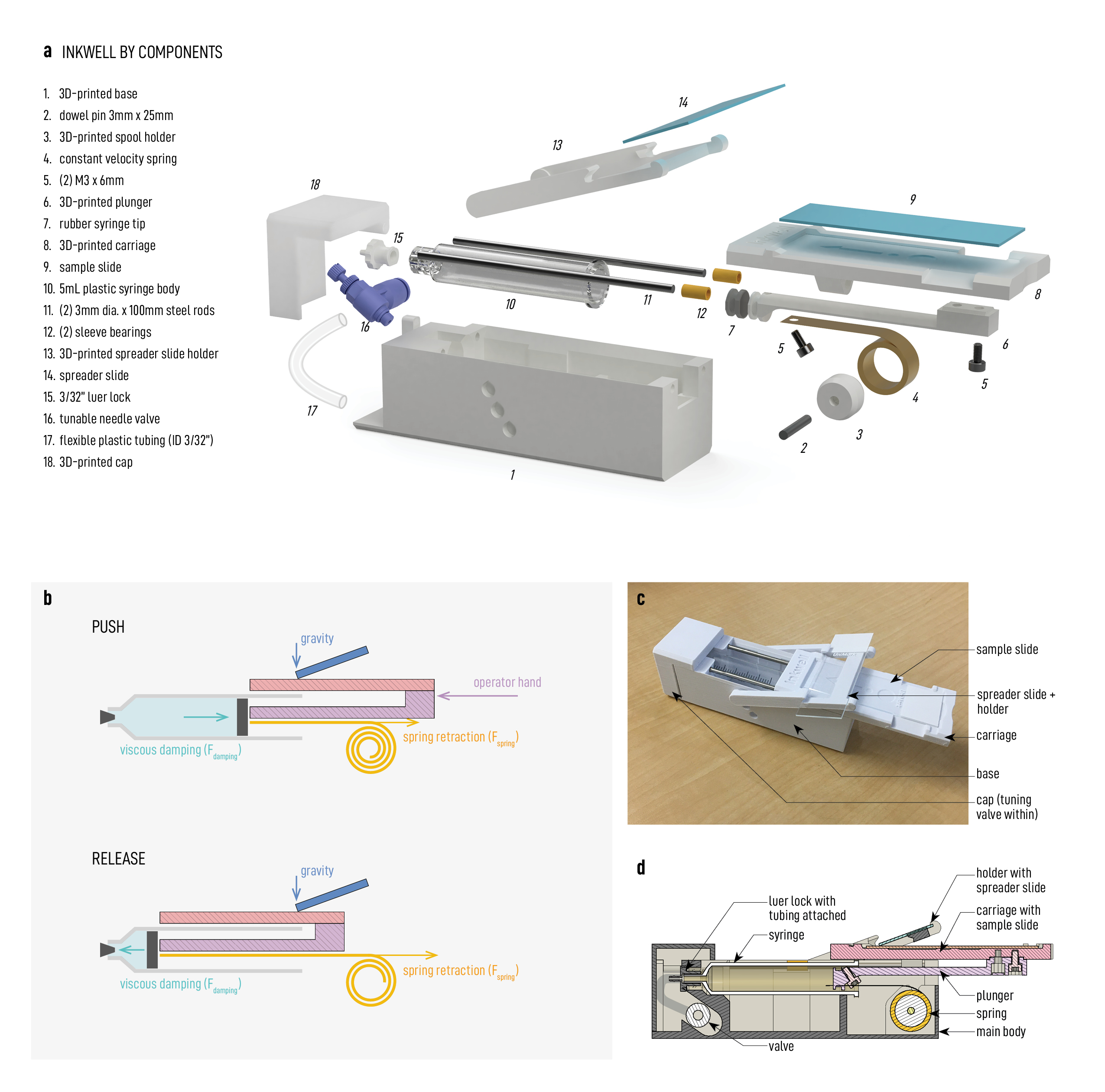}
\caption{\textbf{Inkwell design} [a] Exploded CAD diagram of Inkwell parts consisting of a series of 3D-printed components and off-the-shelf parts. [b] Sectional diagram showing the interactive forces at both the initial push state (user's manual force) and the latter release state (spring force and vacuum pressure). [c] The latest  Inkwell design in lab. [d] CAD sectional diagram of Inkwell showing how the components fit within the unit.}\label{fig4}
\end{figure*}

The final design of Inkwell has three main sub-assemblies: a static base, a carriage that holds the sample slide and translates along the base, and a spreader holder hinged on the base [\ref{fig4}c, d]. The key operation of Inkwell can be summarized as follows. For smooth operation with a constant velocity, we need a carriage and a linear spring coupled to a simple yet reliable dash pot. Here we achieve this by coupling a spiral spring with an air piston regulated by a simple valve as the only control knob for tuning. [\ref{fig4}b]. Once a linear carriage is pushed in and released manually by the user, a constant force spring coupled with an air piston and a regulating valve pushes the carriage back out to its initial position at a constant velocity [\ref{fig4}b]. This outward motion allows the spreader slide to smear the blood across the sample slide. Although initially set to 15-20mm/s, the regulating valve is an adjustable needle valve which lets the user to precisely set the overall smear velocity as needed. The piston is an empty single-use plastic syringe specifically repurposed for our application. Using a syringe and valve combination as an adjustable air damper is novel and dramatically reduces the cost compared to traditional dashpots. See \ref{fig4}a for diagrams and a complete parts list.

Inkwell is easy to make and quick to assemble. It requires only a pair of flush wire cutters and a screwdriver. Teaching somebody how to assemble an Inkwell takes 15 to 20 minutes. Assembly of one unit takes 10 minutes for a newly trained person and 5 minutes for an experienced person. A step-by-step video guide for assembly is provide in supplementary video S6.

\paragraph{Device operation [\ref{fig5}a and video S2]}
Inkwell automates the technically-difficult aspects of producing a good thin smear while still giving the operator control and flexibility over smear thickness and length. We designed and tested Inkwell to operate on various slide brands - we successfully fitted and tested Inkwell with 13 different slide brands spanning 1 mm variation in width and length and 0.25 mm variation in thickness. Therefore any standard slide measuring roughly $76 \times 26 \times 1 \text{ mm}^3$ will be suitable.

% This para is repeated down - so removed. 
%The following are steps to set up Inkwell before use. (1) If the carriage is retracted, unlatch it by lifting the spreader holder and letting it extend out. (2) Set the desired smearing velocity by adjusting the needle valve thumb screw, pushing the carriage in, then observing the velocity at which the carriage self-extends. This may need a few rounds of tuning before the desired velocity is reached. We recommend 15 to 20 mm/s. Our device requires at least two standard $75 \times 25 \times 1 \text{ mm}^3$ microscope glass slides: a sample slide and a spreader slide. 

The operation of Inkwell is shown in \ref{fig5}a. Make sure the carriage is extended out before starting the smearing process:
\begin{enumerate}
    \item Place a clean sample slide in the carriage. Where applicable, the frosted surface (for slide labeling) should be facing up and outward (toward the outer edge of the device).
    \item Insert a spreader slide in the spreader holder, with one of its long (75 mm) edges resting down on the sample slide (make sure the two slides make contact with each other). If the spreader slide has a frosted end, shift the spreader slide so that none of the frosted zone is in contact with the sample slide because that area impedes wetting.
    \item Deposit a sample droplet of blood onto the sample slide over the marked area (~25mm from the end of the sample slide). We recommend 4 µL measured using a pipette or an inverted cup\cite{hopkins2011blood, , incardona2018inverted}.
    \item While holding the base down securely with one hand, push the carriage toward the center of the device with the thumb of the other hand to wind the spring. Take care not to perturb the sample slide. Stop pushing once the drop fully contacts the spreader and spreads across the entire width of the sample slide. Waiting one to five seconds in this position may ensure the droplet is fully spread across the slide before proceeding.
    \item Remove your thumb to allow the carriage to spring back out at the preset velocity. This motion automatically produces a thin smear.
    \item Rotate the spreader along its hinge away from the sample slide and remove the spreader slide, then remove the sample slide.
\end{enumerate}
A single spreader slide has four usable 75-mm edges, therefore can be used to make four separate smears. Between uses, ensure spreader edge is not already contaminated with previous blood samples. The spreader slide can then be cleaned with methanol and reused.

\paragraph{Device performance}
We assessed Inkwell's performance by producing over one hundred whole blood smears in the lab with the single device. In order to test for usability, we asked three lab members, who were newly-trained on Inkwell, to produce thin smears using whole blood from five different samples. \ref{fig5}b shows photos of some of the resulting smears (with an initial volume of 4 µL of whole blood). Regardless of the skill level of the operator, Inkwell can reliably produce high-quality thin smears using whole blood. 

%There were a mix of high-quality and medium-quality smears (due to some stick-slip issues, see Discussion section). 
%However, even in medium-quality smears, the whole smear is covered in a monolayer of densely packed RBCs, as shown on the microscope scans \ref{fig5}c at varying levels of magnification. 

We also compared the smears made with Inkwell against conventional manual smears collected from the field in Uganda and Tanzania. As is generally known, field created smears done manually vary widely in quality and cell density, as is well documented \cite{sori2018external}. For this study we chose two that were representative of the 40 smears collected. Using Octopi, our automated scanning microscope\cite{li2019octopi}, we were able to scan entire smears and stitch them into a gigapixel scan [supplementary scans S1-S4]. We then measured RBC density in each field of view with Cellpose \cite{stringer2021cellpose}. These quantitative measurements demonstrated two advantages of smears made with Inkwell compared to typical hand smears from the field. Firstly, Inkwell produces much more uniform smears, with at least 80 percent of the slide covered in a monolayer. Secondly, Inkwell smears are much denser in the monolayer region, reaching 18,000 RBCs per mm$^2$ in some regions, which is the optimal density before cells start to overlap (see Methods section). \ref{fig6} shows the example scans of Inkwell and manual field smears, with a corresponding heat map of the RBC density distribution across the slide. The gigapixel images for the Inkwell scans show up to 16 million distinguishable RBCs in a single image. To the best of our knowledge, this is the highest number of cells imaged on a single glass slide. There also is an average of up to 16,000 RBCs / mm$^2$, corresponding to 89\% density compared to optimal hexagonal packing across an entire slide [Supplementary scan S1], which is also unprecedented. In total, Inkwell can produce smears with about five times the number of distinguishable RBCs compared to the traditional manual smears we collected. See Table \ref{tab_smears} for a summary of the quantitative comparison.

With Inkwell smears made almost entirely of a monolayer, microscopists and automated diagnostic systems can scan almost anywhere throughout the slide. This contrasts the reliance on the feathered edge zone in conventional smears. Macroscopically, this results in a notable difference between conventional smears which have a tongue-shaped feathered edge and smears made with Inkwell which cover a rectangular area.

\begin{table}[!t]
\caption{Manual Vs. Inkwell Smears: Quality Comparison\label{tab_smears}}%
\begin{tabular*}{\columnwidth}{@{\extracolsep\fill}llll@{\extracolsep\fill}}
\toprule
Metric & Manual Smear (Uganda) & Inkwell Smear (Lab) \\
\midrule
Maximum cell density (cells/mm²) & 11,000 & 18,000 \\
Monolayer area & 20--40 \% & 97--99 \% \\
Average cell density (cells/mm²) & 2,800--4,400 & 13,000--16,000 \\
Total distinguishable cells & 2.8--4.4 million &  13--16 million \\
\bottomrule
\end{tabular*}
\end{table}

\subsection*{Repeatability, robustness, and lifetime testing}
To assess Inkwell's durability over time, we built a motorized stress-testing rig that simulates an operator repeatedly making smears over and over again. Once an Inkwell unit is placed into the rig, a stepper motor and an extended arm repeatedly push and release the carriage to simulate an operator's hand [figS9, Video S4.A]. We test the device for a total of 10,000 cycles. To test performance, thin smears using whole blood before and after 10,000 cycles and the smear qualities between the two are comparable directly [\ref{fig7}a, b]. Even after undergoing 10,000 cycles there was almost no loss of  functionality (in terms of velocity and smoothness of travel) [\ref{fig7}c]. It's worth noting that after thousands of cycles the oil on the syringe barrel and seal gradually migrates away \cite{saad2020cross} which slowly increases friction and reduces the smear velocity (supplementary video S4.B). However this loss in velocity can be compensated by minimally opening the air valve. If, after many more cycles, friction becomes too high, then more oil can easily be added by removing the plunger. This maintenance operation takes about one minute.

\paragraph{Usability study}
A preliminary usability study was conducted in our lab and performed by four lab members to get initial feedback on Inkwell (see Materials and Methods for details). Overall, based upon the post-training survey responses, most participants found Inkwell easy to use and all produced better smears when using Inkwell compared to the conventional method by hand [\ref{fig8}b-d]. Most would consider using Inkwell to produce a large throughput of smears [\ref{fig8}e]. While the current usability study training did not include details on how to properly maintain and repair the device, hence the 50/50 split in confidence level amongst the participants' response to being able to fix the unit [\ref{fig8}f], future training will include such details. Furthermore, a larger usability study with at least 20 field partners and 100 Inkwell units is currently in the planning phase. 

\section*{Discussion}
The overall design of Inkwell over the years has undergone over forty iterations with testing and feedback from users around the world. These iterations range from an all-3D-printed device (including the spring) to motorized systems to ultra-simple compliant mechanisms. However, many of these iterations did not produce the repeatable high-quality smears that we stated as the early goal. Each iteration was tested with pseudo and whole blood and improvements were made in a step-wise manner. Some of the improvements include ways to reduce friction, the number of parts, and the overall assembly time and cost. This iterative process led to [\ref{fig4}] as the final design. 

% \subsection*{Limitations}
One of the main challenges in designing a smearing device such as Inkwell is to ensure carriage motion remains smooth and uniform throughout its life, i.e. thousands to tens of thousands of smears. This goal can be in conflict with lowering manufacturing costs, as cheaper parts tend to be less durable. In particular, the \$ 0.17 syringe we use as an air damper is critical to keep the price of Inkwell low but it is normally  manufactured for single use \cite{fishersyringe}. Keeping the inside surface of the syringe barrel pristine and well-lubricated is ensures a long lifespan for Inkwell. For this,  care should be taken during assembly to not damage or scratch the inside of the syringe and that no contaminants should enter it. In final design, the metal spring, which unspools into the syringe, is mounted with enough clearance with the inside of the barrel in order to avoid scratching the surface. Furthermore, extra silicone oil lubricant is applied to the syringe plug during assembly. See supplementary video S8, where the top device has no extra oil and the bottom device does. Inadequate lubrication can also lead to the carriage jamming halfway or having stick-slip behavior. This leads to non-uniform smears, as shown in \ref{fig5}b (third and fourth smears from the left) which were made with inadequately lubricated devices.

%Without this oil, friction between the plug and barrel is too high, which results in the carriage motion being uneven and degrading over time. 

%In remote, rural environments, some off-the-shelf parts required to build Inkwell may be more difficult to source and need to be replaced by alternatives. For example, the constant force spring could be replaced with a retractable badge reel or a weight attached to a string and pulley. The steel rods on which the carriage translates could be replaced with cut bicycle spokes.

With over 144 million thin smears examined per year \cite{world2022world} for diagnostic purposes. This demonstrates the large need for a low cost device like Inkwell. It is simple to enable multiple smear-making on a single Inkwell (per maneuver) to increase overall smear-making throughput in the field. The current Inkwell device costs 20 dollars in parts - where majority of the cost is driven by 3D printing \ref{tab_BOM}. In large scale using plastic molding, the price can be brought down to as low as 5 dollars per unit. 

%Although the issue of aerosolization (blood droplets as a result of the smear-making) is currently present in other commercially-available products, we will consider additional features to control for aerosolization and potential contamination in the field.

%Additional work will also include design and material choices to allow for easier decontamination protocols. 

To further test Inkwell and it's use in real-world settings, we're currently producing 100 Inkwell devices (using filament 3D-printing, see Supplementary fig S3) to deploy to our field partners and garner feedback on Inkwell's ease of use, ease of maintenance, durability over multiple uses, and quality of smears produced by those working on the ground. 

% 3D printing
The role of 3D printing within the scientific and medical communities has grown and matured over the last decade in terms of precision, quality, and impact. Groups such as OpenFlexture \cite{collins2020robotic} and Open Labware \cite{baden2015open, amann20193d} have compiled various high-quality, open-source designs that are applicable for use in microscopes, micromanipulators, and small-volume liquid handlers. The ability for anyone and anywhere in the world to build quality tools given a simple 3D printer significantly and unprecedentedly improves access to science and research including low resource and remote settings. The design of Inkwell is open-source and will contribute to this large body of work around fabrication and science to improve point-of-care diagnostics including malaria around the world. 

% OpenFlexure: \cite{collins2020robotic}
% Open Labware: \cite{baden2015open}
% \cite{amann20193d}

%%%%%%%%%%%%%%%%%%%%%%%%%%%%%%%%%%%%%%%%%%%%%%%%%%%%%%%%%%%%%%%%%%%%%%%%%%%
\section*{Materials and Methods}

\subsection*{Samples}

Conventional smears were observed in the field in Uganda, Tanzania, and India. Conventional smears in the lab were made using five different samples of human whole blood with EDTA or heparin anticoagulant agents collected from the Stanford Blood Center, and with finger prick blood in the lab. Our motorised parameter sweep study was done with EDTA whole blood from the Stanford Blood Center. Inkwell smears were done in three different settings: (1) in a lab in Uganda with whole blood containing heparin anticoagulant agent, (2) in our Stanford lab with the five samples from the Stanford Blood Center noted above, and (3) in our Stanford lab with fresh finger prick blood.

\subsection*{High-speed imaging of the smearing Process}
To understand the blood-smearing process microscopically, we observed the effects of the smearing action using a custom-built inverted microscope (Squid framework\cite{li2020squid}) in a bio-safety cabinet at 600 fps. In this setup we used an 10x/0.25 objective corrected for 1.1 mm glass thickness, a 75 mm imaging lens as tube lens and an Imaging Source DMK 33UX252 camera. A schematics of the setup can be found in supplementary figure S4 f-g.   

% \subsection*{Experimental Investigation of Optimal Smearing Parameters}

% \paragraph{Requirements for controlled blood smearing experiments}
% One of the main challenges with manual blood smearing is to precisely impart just the right motion and pressure between two slippery pieces of glass moving like a 3D edge slider joint. Indeed this leaves four kinematic degrees of freedom: three for $(x, y)$ planar translation and rotation, and one for spreader tilt angle $phi$.  Only two of these are useful for smearing: $\phi$ and $x$, with the former generally constant throughout the smear. It is left to the operator’s skill to prescribe all these, in addition to controlling downward contact pressure, and the slightest change in any direction has an effect on the final smear. Similarly to Autoheam \cite{mcdermott2022autohaem}, all of this can be automated with a digitally controlled motorized prismatic stage, where set values of $\phi$ are obtained through interchangeable spreader holders. Downward pressure can be controlled e.g. by weighting or spring-loading the spreader.

\subsection*{Design of the motorized characterization setup}
The design of our motorized experimental smearing setup is shown supplementary figure S4 e-g, consisting of 3D-printed and laser-cut parts, a Nema-17 stepper motor with lead screw, and various screws and fasteners. Velocity-controlled prismatic motion is achieved through a 3D-printed carriage actuated by the motor with a TMC2209 motor driver which is controlled by an Arduino Due. There is a magnetic latch on the coupler between the motor and carriage which acts as a mechanical fuse in case the carriage gets pushed too far or is obstructed, and prevents aerosolization or shattered glass slides in the case of mechanical failure. The spreader slide is held in place by an interchangeable slide holder which prescribes the tilt angle (Fig. S4 d). We used 75x25 mm non-beveled slides (3x1 in) and smeared 40--50 mm along the slide, across the entire 25 mm width, leaving room for a potential frosted end for labelling. The spreader slide holder also has integrated flexures with adjustment screws to set the contact pressure with the sample slide. In these parameter-sweep experiments, we used whole human blood with EDTA or heparin anticoagulant. Blood droplet volume was controlled using a pipette.

% \paragraph{Parameter sweeps}
% We performed parameter sweeps for blood volume (2 to 4 µl), velocity (5 to 50 mm/s), and spreader angle (20º to 60º) [Fig. \ref{fig2, figS3}] Each sample was qualitatively inspected for uniformity and symmetry and imaged using an Octopi high-throughput automated microscopy system for quantitative analysis. For tiled scans, shading correction was done using PyBaSic \cite{peng2017basic}.

\subsection*{Inkwell Mechanical Design}
According to the results from the parameter-sweep experiments, to generate a uniform monolayer of red blood cells, we need to drag the glass spreader across the sample slide at constant velocity along a straight line. The easiest way to passively achieve this is to oppose a constant traction force $F_{spring}$ with a viscous damping force $F_{damping}$, much like how terminal velocity is reached by objects falling through a liquid under the action of gravity. By varying the viscous friction coefficient $\alpha$, we can adjust the terminal velocity and therefore the smear thickness. In the case of quadratic friction, this velocity is given by Newton's second law applied to the spreader of mass $m$ subject to these two forces: $m \ddot{x} = F_{spring} - F_{damping} = F_{spring} - \alpha(s) \dot{x}^2$
which yields in steady state: $\dot{x} = \sqrt{\frac{F}{\alpha(s)}}$
where $x$ is the position of the spreader and $s$ formalizes the parameter used to adjust the damping coefficient.

%Todo Jerome: rewrite some of this, it doesn't flow enoguh. Justify why we used a syringe.
%Todo: Which syringes are suitable? Needs large enough piston surface

Inkwell has three main parts: (1) a static base, (2) a carriage that holds the sample slide and translates along the base, and (3) a spreader holder hinged on the base. Structural parts were 3D-printed out of PLA on a Prusa printer (i3 MK3S+). The carriage is actuated by a 500g constant force spring (i.e. a rolled metal strip that recoils when extended \cite{ohtsuki2001analysis}) and dampened by a friction piston. As opposed to using a counterweight, the spring allows for a more compact design and avoids the need to redirect the weight's vertical force. Furthermore, this constant force spring, unlike spiral springs found in e.g. mechanical watches or measuring tapes, at its resting state is recoiled rather than expanded, making the design even more compact. To dampen the spring's motion, we use vacuum pressure in the form of a tunable airflow. Traditionally, this is done by using an expensive dashpot \cite{dashpot}, an ultra-low friction graphite piston designed for smooth and constant dampening. For example commercial smear-making device, Hemaprep, uses a pair of dashpots for the same purpose. However, we found that a plastic syringe \cite{fishersyringe} with extra lubricant, fitted with a luer-lock adaptor and an adjustable needle valve, can provide comparable smooth dampening. Globally, tens of billions of plastic syringes are produced annually and at only a few cents per unit \cite{NPRsyringe}, compared to dashpots which cost as much as \$80 per unit. Using 17 cents syringe significantly reduces the cost of Inkwell.

The needle valve can precisely regulate the amount of airflow and thus fine-tune the steady-state velocity of the carriage. The air reservoir volume between the syringe tip and the needle valve intake is kept as small as possible because the air trapped here expands and acts as a spring when the carriage extends out. As part of the design development, we also tested a variety of tubing (with varying diameters and malleability) that would connect the needle valve to the luer-lock and syringe. The optimal tubing was one that provided sufficient seal, easy to insert into both terminals, and resistant to buckling (to prevent blocking air-flow) (fig. S3 c). 

Gravity and a level mechanism are sufficient to keep contact between the spreader slide and the sample slide. A spring-loaded mechanism to push the spreader down against the sample slide may over-constrain the translation motion and create unnecessary friction. Therefore the spreader holder is loosely hinged on the main body, giving it a little extra play so the edge of the spreader slide can conform to the surface of the sample slide and compensate for any misalignment of the carriage. This is much easier to design and manufacture and will perform much better than a perfectly aligned and constrained system. An over-constrained spreader holder might cause contact loss in case of slight misalignment. We found that the sample slide does not need to be so tightly fixed on the carriage: leaving 0.5 mm to 1 mm of play on each edge of the slide does not affect smear quality.

\paragraph{BOM and cost}
The exact part by part bill of materials for our prototype is given Table \ref{tab_BOM}. Further reduction in per unit cost is easily possible with greater production volume, cheaper off-the-shelf parts, and alternative means of production. For example, injection plastic modling can reduce the cost of 3D printed parts from O(\$10) to O(\$1). 

\begin{table}[!t]
\caption{Inkwell BOM and Cost\label{tab_BOM}}%
\begin{tabular*}{\columnwidth}{@{\extracolsep\fill}llll@{\extracolsep\fill}}
\toprule
Part & Cost at 10 units (USD)  & Cost at 100 units (USD) \\
\midrule
Set of 3D printed parts & \$ 8.50 & \$ 7.50 \\
3mm x 100mm rod (x2) & \$ 0.90 & \$ 0.90 \\
3mm x 25mm rod & \$ 0.20 & \$ 0.20 \\
Igus bearings (x2) & \$ 2.56 & \$ 1.62 \\
Constant force spring & \$ 6.45 & \$ 3.20 \\
Syringe & \$ 0.19 & \$ 0.19 \\
Syringe barb fitting & \$ 0.44 & \$ 0.44 \\
Rubber tube & \$ 0.08 & \$ 0.08 \\
Valve & \$ 0.74 & \$ 0.50 \\
M3 x 6mm screws (x2) & \$ 0.20 & \$ 0.20 \\

\midrule
Total & \$ 20.26 & \$ 14.83 \\
\bottomrule
\end{tabular*}
\end{table}

\section{Long-term durability testing}
To test the long-term durability of Inkwell, we built a stress-testing rig to simulate the usage of Inkwell over thousands of cycles  (fig. S9). The stress-testing rig fits two Inkwell units and can run the simulation in parallel. The rig has a stepper motor and arm that push and release the Inkwell unit(s) per cycle. The rig has adaptable mounting brackets and spacers to fit various versions of Inkwell prototypes. We tested our units 1,000 cycles at a time with an overnight pause between each session to account for mechanical wear and oil drying as seen in normal usage patterns. We found that a unit can withstand 10,000 cycles without any loss in functionality. At close to 20,000 cycles, the device occasionally jams. This can be remedied by loosening the valve and hand-operating the device a few times. If the jamming remains an issue, reapplying oil onto the syringe plunger removes the issue entirely. The predominant source of wear and variability is the gradual oil displacement in the syringe. Other elements like the slide holder have very little wear. Therefore the variation in velocity throughout a single smear (visible macroscopically as bands or multi-gradients within the smear) is sufficient to alert the user to adjust either the valve or reapply oil as necessary to maintain smear quality.

\paragraph{Syringe lubrication}
As noted above, re-lubrication of the syringe plunger may be necessary after thousands of continuous use cycles. We tested mineral oil (universal sewing machine oil) and silicone oil (100 cst viscocity) to smoothen the carriage motion. For this test, we applied the experimented oil on the syringe plunger during assembly (about 5-10 µL) and cycled the device until there was a noticeable increase in friction. We found that mineral oil dries relatively quickly and results in high stiction, whereas silicone oil last for tens of thousands of cycles and does not dry as easily. We also observed oil being wiped off whenever the plunger is removed for maintenance, therefore to prevent stiction, more oil needs to be reapplied each time the plunger is removed. Additionally, if the plug/rubber tip is pushed in all the way until it butts up against the inner tip of the syringe, the plunger can get stuck from  oil making an internal wet seal and that excess oil by depositing into the inside tip of the syringe. Our current design has a built-in mechanical stop that prevents this from happening.

\paragraph{Usability study}
Since Inkwell was designed primarily for untrained, unskilled individuals, the four participants in a preliminary usability test were selected based on their inexperience with making thin blood smears [\ref{fig8}a]. The usability study was done in two identical sessions (two participants each) where participants were minimally trained on how thin blood smears are conventionally made and shown how to operate Inkwell to produce thin blood smears. Examples of "good" and "bad" smears were presented to each participant as part of the training. To eliminate the need for additional safety training on handling blood, a phantom blood solution (designed to mimic the rheological properties of human whole blood and was prepared according to IS 5405:1980) was used. Each participant was allowed to practice smear-making on roughly three to five slides, after which they are asked to produce five thin smears by hand and five thin smears using Inkwell. The initial volume of phantom blood was set at 4 µL per smear. After producing the ten smears, each participant was given a short questionnaire to fill out [\ref{fig8}b-f]. The questions include rating their ability to produce smears using Inkwell and the quality of smears by Inkwell in comparison to the conventional manual method. Additional questions ask participants to briefly describe any shortcomings or misunderstandings of the instructions and/or of the device, and whether they would consider using Inkwell for future fieldwork or not, and why.

\section*{Acknowledgements}
We thank the following individuals for participating in the Inkwell usability study: Ethan Li, Hope Leng, Grace Zhong, Ray Chang, and Jijumon A.S. We thank Ipsita Pal Bhowmick for discussions and images used in figure 1. We thank all members of the PrakashLab for fruitful discussions and comments. This work was funded by Gates Foundation Award (M.P), Schmidt Futures Innovation Fellowship (M.P) and Moore Foundation Research Grant (M.P.) and NSF CCC (DBI1548297 (M.P.)). M.P., H.L. and A.K. identified the research problem. M.P., A.K. and D.A. developed the initial prototypes and experiments. J.N. and A.K. developed prototypes and implemented the final Inkwell device with input from all authors. J.N., H.L. and A.K. set up the motorized smearing device for characterizations. H.L. set up the apparatus for high-speed imaging and the Octopi microscope for scanning the blood smears. J.N. and A.K. performed data collection. A.K. designed the usability studies. J.P. validated an earlier prototype in field settings. J.N., H.L., A.K., and M.P. wrote the manuscript with feedback from all authors. 

\section*{Figures and Captions}

\begin{figure*}[!h]%
\centering
\includegraphics[width=500pt]{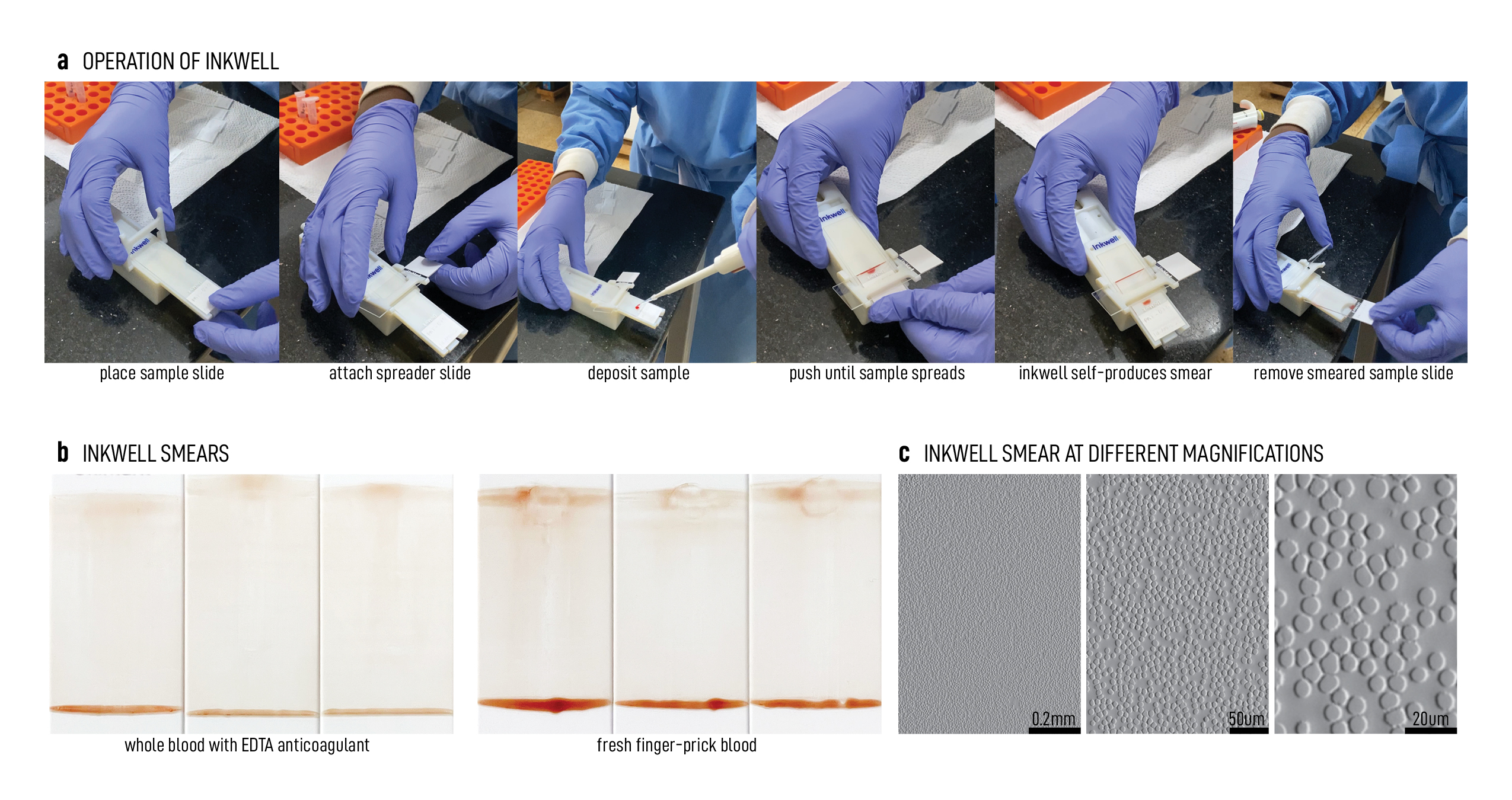}
\caption{\textbf{Functionality of Inkwell} [a] Step-by-step how-to photos of a technician using Inkwell to produce a thin smear. [b] Comparative array of thin smears produced using Inkwell at Stanford by a minimally-trained lab member using whole blood containing EDTA as anticoagulant versus fresh finger-prick blood. [c] Microscopic scan of a thin smear produced by Inkwell at 3 different magnifications showing the uniformity of density of monolayer of red blood cells.}
\label{fig5}
\end{figure*}

\begin{figure*}[!h]%
\centering
\includegraphics[width=500pt]{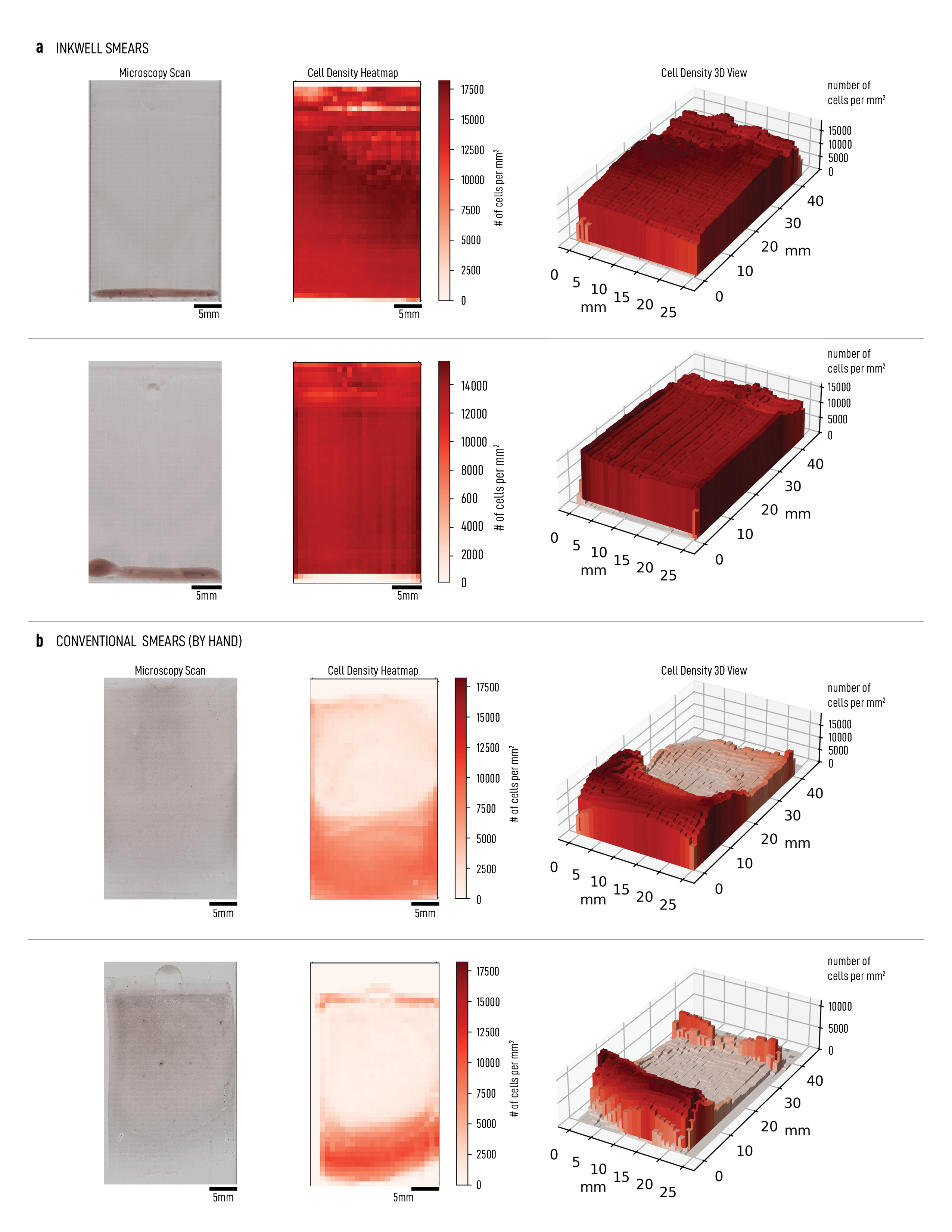}
\caption{\textbf{Evaluation of smears made with Inkwell}
[a] Scans along with 2D heatmaps and a 3D representations of predicted per-field-of-view cell density of two selected thin-smears produced using Inkwell. [b] As a comparison, similar scans and and cell density representations were done for two conventional smears made by hand.}\label{fig6}
\end{figure*}

\begin{figure*}[!h]%
\centering
\includegraphics[width=500pt]{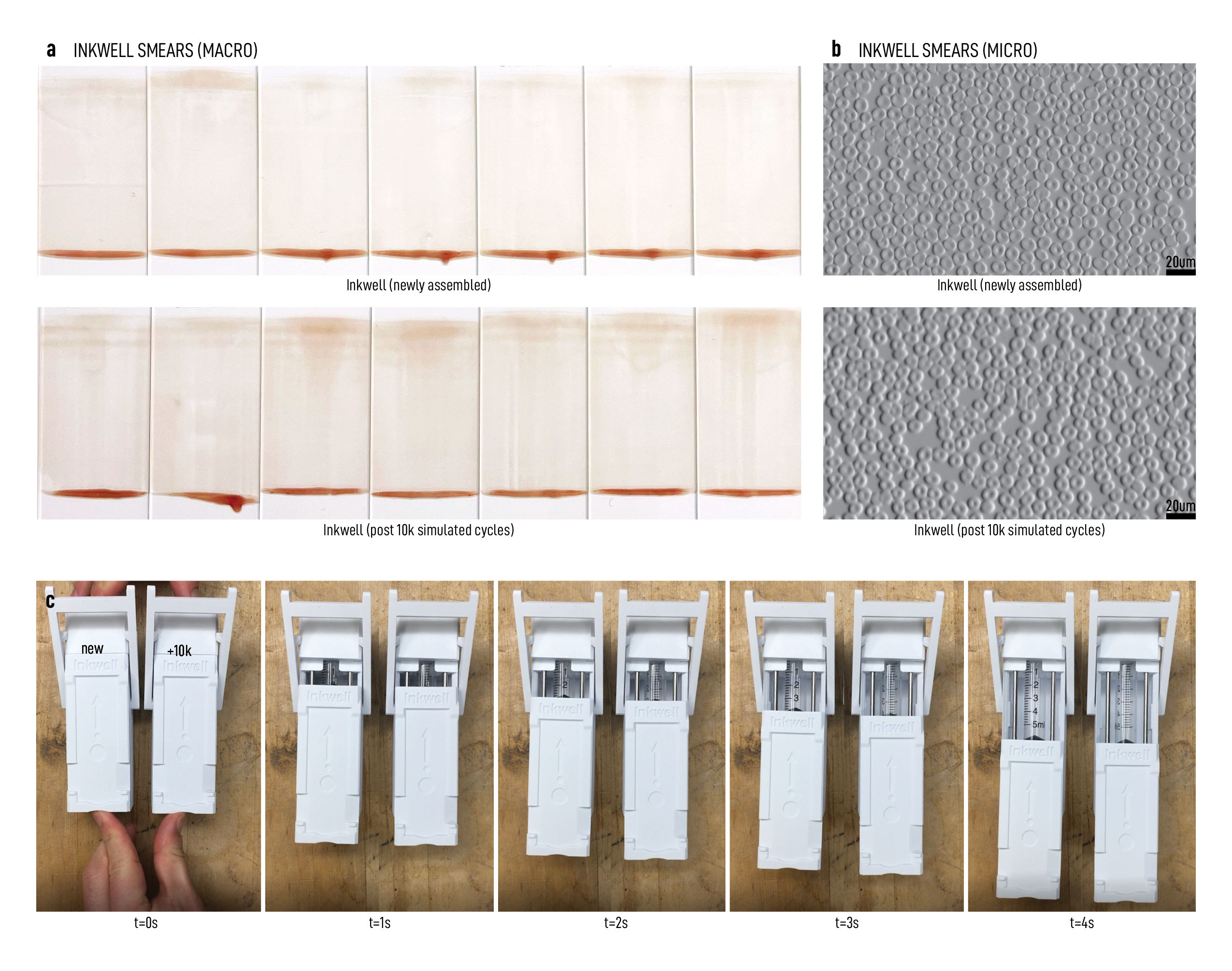}
\caption{\textbf{Inkwell tested 10,000 cycles} 
To assess the robustness over time, Inkwell was subjected to multiple push-release cycles with the aid of a motorized system (see Materials and Methods). [a] Thin blood smears produced using two different Inkwell units, one newly assembled and another which had undergone 10,000 simulated cycles, were of comparable quality to each other. [b] Scans of the thin blood smears from [a] show comparable RBC densities. [c] Observed velocities across the two Inkwell units (left new, right after 10,000 test cycle run) show comparable motions of the carriage (a readout of uniform and consistent velocity) without reduction in smoothness. 
}\label{fig7}
\end{figure*}

\begin{figure*}[!h]%
\centering
\includegraphics[width=500pt]{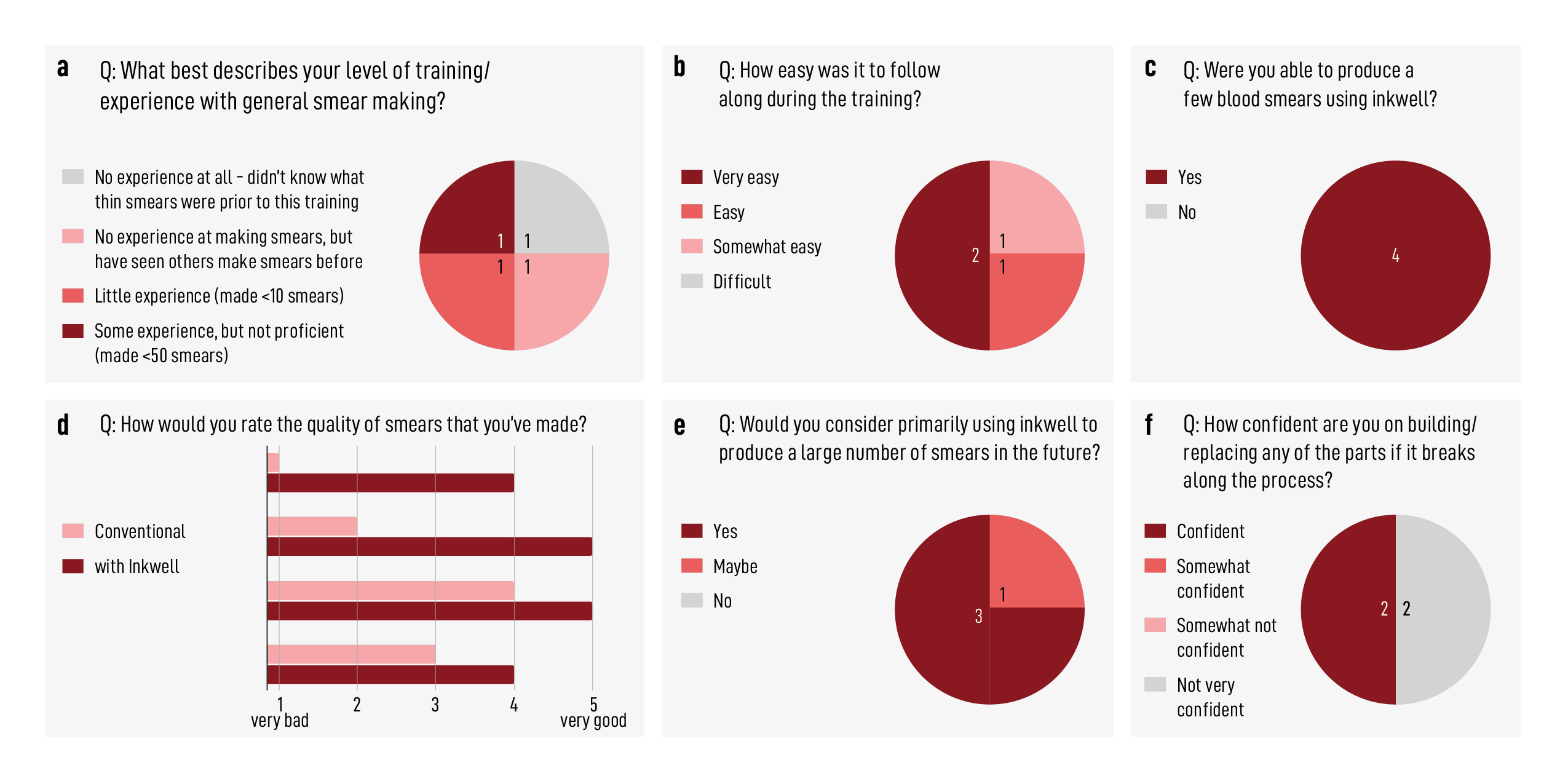}
\caption{\textbf{Inkwell usability survey}
Questions and answers from the usability test survey that each volunteer was asked to fill out post the training session. [a] All four volunteers were selected for having little to no experience in making thin smears - the target user group for Inkwell. [b] Most participants noted that Inkwell was easy learn to use following the brief training. [c] All participants were able to produce smears using Inkwell. [d] All participants rated the smears produced using Inkwell exceed the quality of those produced conventionally by hand. [e] Most participants would consider using Inkwell to produce smears in the future, especially for high throughput. [f] Half of the participants were confident and the other half not in fixing or replacing any potentially broken elements of Inkwell. However, the training only demonstrated how to tune the velocity using the valve knob and did not cover how to fix broken parts.}
\label{fig8}
\end{figure*}

% \bibliographystyle{naturemag}
% \bibliography{main_bib}

\FloatBarrier
\newpage
\bibliographystyle{naturemag}
\bibliography{bibliography.bib}

\begin{thebibliography}{10}
\expandafter\ifx\csname url\endcsname\relax
  \def\url#1{\texttt{#1}}\fi
\expandafter\ifx\csname urlprefix\endcsname\relax\def\urlprefix{URL }\fi
\providecommand{\bibinfo}[2]{#2}
\providecommand{\eprint}[2][]{\url{#2}}

\bibitem{world2016malaria}
\bibinfo{author}{Organization, W.~H.}
\newblock \emph{\bibinfo{title}{Malaria microscopy quality assurance
  manual-version 2}} (\bibinfo{publisher}{World Health Organization},
  \bibinfo{year}{2016}).

\bibitem{sori2018external}
\bibinfo{author}{Sori, G.}, \bibinfo{author}{Zewdie, O.},
  \bibinfo{author}{Tadele, G.} \& \bibinfo{author}{Samuel, A.}
\newblock \bibinfo{title}{External quality assessment of malaria microscopy
  diagnosis in selected health facilities in western oromia, ethiopia}.
\newblock \emph{\bibinfo{journal}{Malaria journal}}
  \textbf{\bibinfo{volume}{17}}, \bibinfo{pages}{1--7} (\bibinfo{year}{2018}).

\bibitem{mutabazi2021assessment}
\bibinfo{author}{Mutabazi, T.} \emph{et~al.}
\newblock \bibinfo{title}{Assessment of the accuracy of malaria microscopy in
  private health facilities in entebbe municipality, uganda: a cross-sectional
  study}.
\newblock \emph{\bibinfo{journal}{Malaria Journal}}
  \textbf{\bibinfo{volume}{20}}, \bibinfo{pages}{1--9} (\bibinfo{year}{2021}).

\bibitem{guerin2002malaria}
\bibinfo{author}{Guerin, P.~J.} \emph{et~al.}
\newblock \bibinfo{title}{Malaria: current status of control, diagnosis,
  treatment, and a proposed agenda for research and development}.
\newblock \emph{\bibinfo{journal}{The Lancet Infectious Diseases}}
  \textbf{\bibinfo{volume}{2}}, \bibinfo{pages}{564--573}
  (\bibinfo{year}{2002}).

\bibitem{world2010basic_tutor}
\bibinfo{author}{Organization, W.~H.} \& \bibinfo{author}{for Disease~Control,
  C.}
\newblock \emph{\bibinfo{title}{Basic malaria microscopy: tutor's guide}}
  (\bibinfo{publisher}{World Health Organization}, \bibinfo{year}{2010}).

\bibitem{li2019octopi}
\bibinfo{author}{Li, H.}, \bibinfo{author}{Soto-Montoya, H.},
  \bibinfo{author}{Voisin, M.}, \bibinfo{author}{Valenzuela, L.~F.} \&
  \bibinfo{author}{Prakash, M.}
\newblock \bibinfo{title}{Octopi: Open configurable high-throughput imaging
  platform for infectious disease diagnosis in the field}.
\newblock \emph{\bibinfo{journal}{BioRxiv}} \bibinfo{pages}{684423}
  (\bibinfo{year}{2019}).

\bibitem{das2022field}
\bibinfo{author}{Das, D.} \emph{et~al.}
\newblock \bibinfo{title}{Field evaluation of the diagnostic performance of
  easyscan go: a digital malaria microscopy device based on machine-learning}.
\newblock \emph{\bibinfo{journal}{Malaria Journal}}
  \textbf{\bibinfo{volume}{21}}, \bibinfo{pages}{122} (\bibinfo{year}{2022}).

\bibitem{maguire2006production}
\bibinfo{author}{Maguire, J.~D.} \emph{et~al.}
\newblock \bibinfo{title}{Production and validation of durable, high quality
  standardized malaria microscopy slides for teaching, testing and quality
  assurance during an era of declining diagnostic proficiency}.
\newblock \emph{\bibinfo{journal}{Malaria Journal}}
  \textbf{\bibinfo{volume}{5}}, \bibinfo{pages}{1--8} (\bibinfo{year}{2006}).

\bibitem{kahama2011low}
\bibinfo{author}{Kahama-Maro, J.}, \bibinfo{author}{D'Acremont, V.},
  \bibinfo{author}{Mtasiwa, D.}, \bibinfo{author}{Genton, B.} \&
  \bibinfo{author}{Lengeler, C.}
\newblock \bibinfo{title}{Low quality of routine microscopy for malaria at
  different levels of the health system in dar es salaam}.
\newblock \emph{\bibinfo{journal}{Malaria Journal}}
  \textbf{\bibinfo{volume}{10}}, \bibinfo{pages}{1--10} (\bibinfo{year}{2011}).

\bibitem{harchut2013over}
\bibinfo{author}{Harchut, K.} \emph{et~al.}
\newblock \bibinfo{title}{Over-diagnosis of malaria by microscopy in the
  kilombero valley, southern tanzania: an evaluation of the utility and
  cost-effectiveness of rapid diagnostic tests}.
\newblock \emph{\bibinfo{journal}{Malaria journal}}
  \textbf{\bibinfo{volume}{12}}, \bibinfo{pages}{1--9} (\bibinfo{year}{2013}).

\bibitem{ngasala2019evaluation}
\bibinfo{author}{Ngasala, B.} \& \bibinfo{author}{Bushukatale, S.}
\newblock \bibinfo{title}{Evaluation of malaria microscopy diagnostic
  performance at private health facilities in tanzania}.
\newblock \emph{\bibinfo{journal}{Malaria journal}}
  \textbf{\bibinfo{volume}{18}}, \bibinfo{pages}{1--7} (\bibinfo{year}{2019}).

\bibitem{CDC_thinsmear}
\bibinfo{author}{for Disease~Control, C.} \& \bibinfo{author}{Prevention}.
\newblock \bibinfo{title}{Specimen processing} (\bibinfo{year}{2020}).
\newblock
  \urlprefix\url{https://www.cdc.gov/dpdx/diagnosticprocedures/blood/specimenproc.html}.

\bibitem{world2010basic_learner}
\bibinfo{author}{Organization, W.~H.} \emph{et~al.}
\newblock \emph{\bibinfo{title}{Basic malaria microscopy: Part I. Learner's
  guide.}}
\newblock \bibinfo{number}{Ed. 2} (\bibinfo{publisher}{World Health
  Organization}, \bibinfo{year}{2010}).

\bibitem{world2016collection}
\bibinfo{author}{Organization, W.~H.} \emph{et~al.}
\newblock \bibinfo{title}{Collection of finger-prick blood and preparation of
  thick and thin blood films}.
\newblock \bibinfo{type}{Tech. Rep.}, \bibinfo{institution}{World Health
  Organization} (\bibinfo{year}{2016}).

\bibitem{adewoyin2014peripheral}
\bibinfo{author}{Adewoyin, A.}
\newblock \bibinfo{title}{Peripheral blood film-a review}.
\newblock \emph{\bibinfo{journal}{Annals of Ibadan postgraduate medicine}}
  \textbf{\bibinfo{volume}{12}}, \bibinfo{pages}{71--79}
  (\bibinfo{year}{2014}).

\bibitem{moura2014impact}
\bibinfo{author}{Moura, S.} \emph{et~al.}
\newblock \bibinfo{title}{Impact of a training course on the quality of malaria
  diagnosis by microscopy in angola}.
\newblock \emph{\bibinfo{journal}{Malaria Journal}}
  \textbf{\bibinfo{volume}{13}}, \bibinfo{pages}{1--7} (\bibinfo{year}{2014}).

\bibitem{perfectsmear}
\bibinfo{author}{AmScope}.
\newblock \bibinfo{title}{Globe scientific diamond perfect smear blood smearing
  tool}.
\newblock
  \urlprefix\url{https://amscope.com/collections/globe-scientific/products/c-vgsc-1300}.

\bibitem{saad2020cross}
\bibinfo{author}{Saad~Albichr, I.} \emph{et~al.}
\newblock \bibinfo{title}{Cross-evaluation of five slidemakers and three
  automated image analysis systems: The pitfalls of automation?}
\newblock \emph{\bibinfo{journal}{International Journal of Laboratory
  Hematology}} \textbf{\bibinfo{volume}{42}}, \bibinfo{pages}{573--580}
  (\bibinfo{year}{2020}).

\bibitem{mcdermott2022autohaem}
\bibinfo{author}{McDermott, S.} \emph{et~al.}
\newblock \bibinfo{title}{autohaem: 3d printed devices for automated
  preparation of blood smears}.
\newblock \emph{\bibinfo{journal}{Review of Scientific Instruments}}
  \textbf{\bibinfo{volume}{93}}, \bibinfo{pages}{014104}
  (\bibinfo{year}{2022}).

\bibitem{sysmex}
\bibinfo{author}{Sysmex}.
\newblock \bibinfo{title}{Seed haematology, sysmex educational enhancement and
  development}.
\newblock \bibinfo{type}{Tech. Rep.}, \bibinfo{institution}{Sysmex}
  (\bibinfo{year}{2013}).

\bibitem{sysmex2}
\bibinfo{author}{Sysmex}.
\newblock \bibinfo{title}{Ral stainer}.
\newblock
  \urlprefix\url{https://www.sysmex-mea.com/products/products-detail/ral-stainer.html}.

\bibitem{britten1993simple}
\bibinfo{author}{Britten, J.~A.}
\newblock \bibinfo{title}{A simple theory for the entrained film thickness
  during meniscus coating}.
\newblock \emph{\bibinfo{journal}{Chemical Engineering Communications}}
  \textbf{\bibinfo{volume}{120}}, \bibinfo{pages}{59--71}
  (\bibinfo{year}{1993}).

\bibitem{le2009convective}
\bibinfo{author}{Le~Berre, M.}, \bibinfo{author}{Chen, Y.} \&
  \bibinfo{author}{Baigl, D.}
\newblock \bibinfo{title}{From convective assembly to landau- levich deposition
  of multilayered phospholipid films of controlled thickness}.
\newblock \emph{\bibinfo{journal}{Langmuir}} \textbf{\bibinfo{volume}{25}},
  \bibinfo{pages}{2554--2557} (\bibinfo{year}{2009}).

\bibitem{reznik2009entrainment}
\bibinfo{author}{Reznik, S.}, \bibinfo{author}{Salalha, W.},
  \bibinfo{author}{Sorek, Y.}, \bibinfo{author}{Avramov, D.} \&
  \bibinfo{author}{Zussman, E.}
\newblock \bibinfo{title}{Entrainment of a film on a surface from the meniscus
  of a liquid wedge during coating}.
\newblock \emph{\bibinfo{journal}{Physics of Fluids}}
  \textbf{\bibinfo{volume}{21}}, \bibinfo{pages}{102001}
  (\bibinfo{year}{2009}).

\bibitem{mayer2012landau}
\bibinfo{author}{Mayer, H.} \& \bibinfo{author}{Krechetnikov, R.}
\newblock \bibinfo{title}{Landau-levich flow visualization: Revealing the flow
  topology responsible for the film thickening phenomena}.
\newblock \emph{\bibinfo{journal}{Physics of Fluids}}
  \textbf{\bibinfo{volume}{24}}, \bibinfo{pages}{052103}
  (\bibinfo{year}{2012}).

\bibitem{thurston1972viscoelasticity}
\bibinfo{author}{Thurston, G.~B.}
\newblock \bibinfo{title}{Viscoelasticity of human blood}.
\newblock \emph{\bibinfo{journal}{Biophysical journal}}
  \textbf{\bibinfo{volume}{12}}, \bibinfo{pages}{1205--1217}
  (\bibinfo{year}{1972}).

\bibitem{errill1969rheology}
\bibinfo{author}{Errill, E.}
\newblock \bibinfo{title}{Rheology of blood}.
\newblock \emph{\bibinfo{journal}{Physiological reviews}}
  \textbf{\bibinfo{volume}{49}}, \bibinfo{pages}{863--888}
  (\bibinfo{year}{1969}).

\bibitem{doumenc2016modeling}
\bibinfo{author}{Doumenc, F.}, \bibinfo{author}{Salmon, J.-B.} \&
  \bibinfo{author}{Guerrier, B.}
\newblock \bibinfo{title}{Modeling flow coating of colloidal dispersions in the
  evaporative regime: prediction of deposit thickness}.
\newblock \emph{\bibinfo{journal}{Langmuir}} \textbf{\bibinfo{volume}{32}},
  \bibinfo{pages}{13657--13668} (\bibinfo{year}{2016}).

\bibitem{gutenev2003liquid}
\bibinfo{author}{Gutenev, P.}, \bibinfo{author}{Pyatnitskii, A.} \&
  \bibinfo{author}{Klimova, N.}
\newblock \bibinfo{title}{Liquid entrainment from the meniscus of a liquid
  wedge by a moving horizontal plate}.
\newblock \emph{\bibinfo{journal}{Colloid Journal}}
  \textbf{\bibinfo{volume}{65}}, \bibinfo{pages}{301--304}
  (\bibinfo{year}{2003}).

\bibitem{cheng2004complete}
\bibinfo{author}{Cheng, C. K.-W.}, \bibinfo{author}{Chan, J.},
  \bibinfo{author}{Cembrowski, G.~S.} \& \bibinfo{author}{van Assendelft,
  O.~W.}
\newblock \bibinfo{title}{Complete blood count reference interval diagrams
  derived from nhanes iii: stratification by age, sex, and race}.
\newblock \emph{\bibinfo{journal}{Laboratory Hematology}}
  \textbf{\bibinfo{volume}{10}}, \bibinfo{pages}{42--53}
  (\bibinfo{year}{2004}).

\bibitem{hopkins2011blood}
\bibinfo{author}{Hopkins, H.} \emph{et~al.}
\newblock \bibinfo{title}{Blood transfer devices for malaria rapid diagnostic
  tests: evaluation of accuracy, safety and ease of use}.
\newblock \emph{\bibinfo{journal}{Malaria journal}}
  \textbf{\bibinfo{volume}{10}}, \bibinfo{pages}{1--9} (\bibinfo{year}{2011}).

\bibitem{incardona2018inverted}
\bibinfo{author}{Incardona, S.} \emph{et~al.}
\newblock \bibinfo{title}{The inverted cup device for blood transfer on malaria
  rdts: ease of use, acceptability and safety in routine use by health workers
  in nigeria}.
\newblock \emph{\bibinfo{journal}{Malaria Journal}}
  \textbf{\bibinfo{volume}{17}}, \bibinfo{pages}{1--8} (\bibinfo{year}{2018}).

\bibitem{stringer2021cellpose}
\bibinfo{author}{Stringer, C.}, \bibinfo{author}{Wang, T.},
  \bibinfo{author}{Michaelos, M.} \& \bibinfo{author}{Pachitariu, M.}
\newblock \bibinfo{title}{Cellpose: a generalist algorithm for cellular
  segmentation}.
\newblock \emph{\bibinfo{journal}{Nature methods}}
  \textbf{\bibinfo{volume}{18}}, \bibinfo{pages}{100--106}
  (\bibinfo{year}{2021}).

\bibitem{fishersyringe}
\bibinfo{author}{Scientific, F.}
\newblock \bibinfo{title}{Fisherbrand sterile syringes for single use}.
\newblock
  \urlprefix\url{https://www.fishersci.com/shop/products/sterile-syringes-single-use-12/14955458}.

\bibitem{world2022world}
\bibinfo{author}{Organization, W.~H.} \emph{et~al.}
\newblock \emph{\bibinfo{title}{World malaria report 2022}}
  (\bibinfo{publisher}{World Health Organization}, \bibinfo{year}{2022}).

\bibitem{collins2020robotic}
\bibinfo{author}{Collins, J.~T.} \emph{et~al.}
\newblock \bibinfo{title}{Robotic microscopy for everyone: the openflexure
  microscope}.
\newblock \emph{\bibinfo{journal}{Biomedical Optics Express}}
  \textbf{\bibinfo{volume}{11}}, \bibinfo{pages}{2447--2460}
  (\bibinfo{year}{2020}).

\bibitem{baden2015open}
\bibinfo{author}{Baden, T.} \emph{et~al.}
\newblock \bibinfo{title}{Open labware: 3-d printing your own lab equipment}.
\newblock \emph{\bibinfo{journal}{PLoS biology}} \textbf{\bibinfo{volume}{13}},
  \bibinfo{pages}{e1002086} (\bibinfo{year}{2015}).

\bibitem{amann20193d}
\bibinfo{author}{Amann, S.}, \bibinfo{author}{Witzleben, M.~v.} \&
  \bibinfo{author}{Breuer, S.}
\newblock \bibinfo{title}{3d-printable portable open-source platform for
  low-cost lens-less holographic cellular imaging}.
\newblock \emph{\bibinfo{journal}{Scientific reports}}
  \textbf{\bibinfo{volume}{9}}, \bibinfo{pages}{1--10} (\bibinfo{year}{2019}).

\bibitem{li2020squid}
\bibinfo{author}{Li, H.} \emph{et~al.}
\newblock \bibinfo{title}{Squid: simplifying quantitative imaging platform
  development and deployment}.
\newblock \emph{\bibinfo{journal}{bioRxiv}} \bibinfo{pages}{2020--12}
  (\bibinfo{year}{2020}).

\bibitem{ohtsuki2001analysis}
\bibinfo{author}{Ohtsuki, A.}, \bibinfo{author}{Ohshima, S.} \&
  \bibinfo{author}{Itoh, D.}
\newblock \bibinfo{title}{Analysis on characteristics of a c-shaped
  constant-force spring with a guide}.
\newblock \emph{\bibinfo{journal}{JSME International Journal Series C
  Mechanical Systems, Machine Elements and Manufacturing}}
  \textbf{\bibinfo{volume}{44}}, \bibinfo{pages}{494--499}
  (\bibinfo{year}{2001}).

\bibitem{dashpot}
\bibinfo{author}{Corp, A.}
\newblock \bibinfo{title}{Airpot}.
\newblock \urlprefix\url{https://www.airpot.com/free-samples-push-dashpot/}.

\bibitem{NPRsyringe}
\bibinfo{author}{Radio, N.~P.}
\newblock \bibinfo{title}{A looming challenge in the vaccination campaign:
  Syringe shortages} (\bibinfo{year}{2021}).
\newblock
  \urlprefix\url{https://www.npr.org/2021/07/25/1020488395/a-looming-challenge-in-the-vaccination-campaign-syringe-shortages}.

\end{thebibliography}

%\beginsupplement
\newpage
%-----------------------------------------------------------

\end{document}

\title{Inkwell: Design and Validation of a Low-Cost Electricity-Free Device for Automated Thin Smearing of Whole Blood: Supplementary Information}

\author[1,$\dagger$]{Jerome Nowak}
\author[2,$\dagger$]{Anesta Kothari}
\author[3]{Hongquan Li}
\author[4]{Jassi Pannu}
\author[5]{Dani Algazi}
\author[2,6,7,8,*]{Manu Prakash}

\affil[1]{Department of Mechanical Engineering}
\affil[2]{Department of Bioengineering}
\affil[3]{Department of Electrical Engineering}
\affil[4]{School of Medicine}
\affil[5]{Department of Computer Science}
\affil[6]{Department of Biology}
\affil[7]{Woods Institute of the Environment}
\affil[8]{Center for Innovation in Global Health}
\affil[]{Stanford University, Stanford, California, USA}
\affil[$\dagger$]{co-first author}
\affil[*]{To whom correspondence should be addressed: manup@stanford.edu}
\date{\today}

\maketitle
\newpage

\makeatletter
\renewcommand{\thefigure}{Fig. S\arabic{figure}}
\renewcommand{\figurename}{}
\renewcommand{\themovie}{Movie S\arabic{movie}}

\makeatother
\setcounter{figure}{0}
\setcounter{section}{0}
\setcounter{table}{0}
\setcounter{page}{1}

\newpage

\tableofcontents
%=========================================================================================================
\newpage

\section{List of Supplementary Materials}
\begin{itemize}
    \item Figs. \ref{figS1} to \ref{figS9}. Supplementary figures.
    \item Table S1. Cost estimate for Inkwell.
    \item Supplementary videos S1 to S7 ( \href{https://drive.google.com/drive/folders/1MlT_pgtp0PvJhoWDKAJIDc6DsJXu7Qbr?usp=sharing}{see Google Drive folder}) %[Todo: replace with final location])
        \begin{itemize}
            \item Videos S1 A and B: Slow motion microscopy imagery of human blood smeared at various velocities, with 2µl (A) and 3µl (B) blood volumes.
            \item Videos S2 A and B: Operation of Inkwell producing a thin blood smear with finger prick blood (A) and with EDTA blood (B). Note: Video S2B was made with an older iteration of Inkwell.
            \item Video S3 A and B: Pan and zoom into a microscope scan of a manual blood smear (A) and an Inkwell blood smear. The scans were made with our in-house Octopi microscope.
            \item Video S4: Inkwell step-by-step assembly instructions.
            \item Videos S5 A and B: (A) automated stress testing rig running two Inkwell through smearing cycles. (B) comparison between a brand new Inkwell unit (bottom) and one that underwent 10,000 cycles (top), where both units were calibrated at the exact same velocity prior to stress testing.
            \item Video S6: Comparison between two Inkwell units with the syringe not lubricated (top) and lubricated (bottom). The air valve is fully open on both units so the predominant source of friction is from the syringe.
            \item Videos S7 A and B: Early prototypes of Inkwell using a dashpot instead of a syringe for air damping.
        \end{itemize}
    \item Supplementary scans S1 and S2
        \begin{itemize}
            \item \href{https://storage.googleapis.com/inkwell/6XnKIqVL.html}{Scan S1}, Inkwell smear from the lab
            %\item \href{http://www.gigapan.org/gigapans/b021192247300696a93bef7db7b8cdba}{Scan S2}, Inkwell smear from the lab
            \item \href{https://storage.googleapis.com/inkwell/Xj9tT0AC.html}{Scan S2}, conventional manual smear from Uganda
            %\item \href{http://www.gigapan.org/gigapans/230609}{Scan S4}, conventional manual smear from Uganda
        \end{itemize}
\end{itemize}

\section{Smear Data Analysis}

We imaged our thin smears with our Octopi automated microscope in bright field \cite{li2019octopi}, which outputs $0.93 \times 0.93 \text{ mm}^2$ fields of view as RGB images measuring and $3000 \times 3000 \text{ pixels}$. The surface area was measured using a calibration slide with a reticle calibrating ruler. To count the number of individually distinguishable RBCs per unit surface (that is to say, the number of cells in the monolayer), as presented Fig. \ref{fig6} and Table \ref{tab_smears}, we used Cellpose \cite{stringer2021cellpose}. First, we processed the green channel of each field of view through the \textit{cyto} model set to a 30 pixel diameter [Fig. \ref{figS4}]. Second, we counted the number of cell masks which Cellpose outputted. Since Cellpose does not detect dense clusters of overlapping and merged cells, this gave us a good estimation of the number of monolayered cells. No pre-training was required. Our code is based on the provided sample notebook \cite{cellpose_notebook}, though more recent versions have since been published \cite{stringer2022cellpose}.

The microscope scans included in our figures were post-processed by removing the background illumination, where each pixel value was normalised by the baseline illumination for an empty glass slide. This baseline illumination was computed from the median over nine locations on the slide to remove any dust or debris. Each location was averaged over ten images to minimise thermal noise in the sensor and light source. For grayscale figures, we then computed the differential phase contrast between the green channels of two images illuminated from opposite sides (right and left). For colour images, we then corrected the white balance using the MATLAB \texttt{chromadapt} function by selecting a pixel from an empty region of the slide.

\newpage

\section{Supplementary Figures}
%-----------------------------------------------------------

% Supplementary figures

\begin{figure*}[!h]%
\centering
\includegraphics[width=500pt]{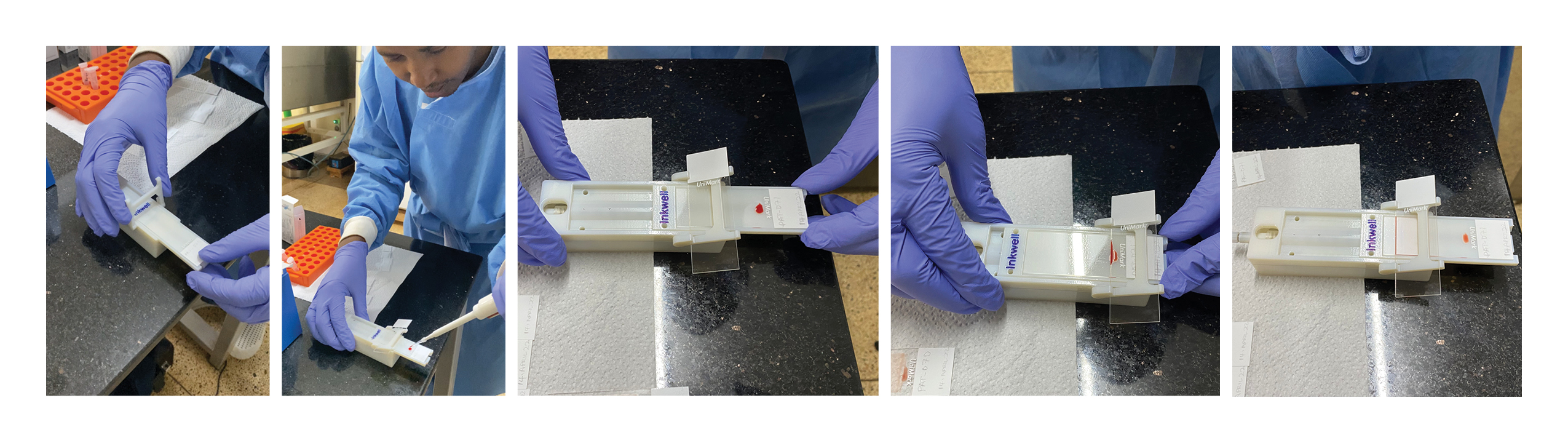}
\caption{\textbf{Inkwell used abroad}
Inkwell being used by a technician in a lab in Uganda.}\label{figS1}
\end{figure*}

\clearpage

\begin{figure*}[!h]%
\centering
\includegraphics[width=455pt]{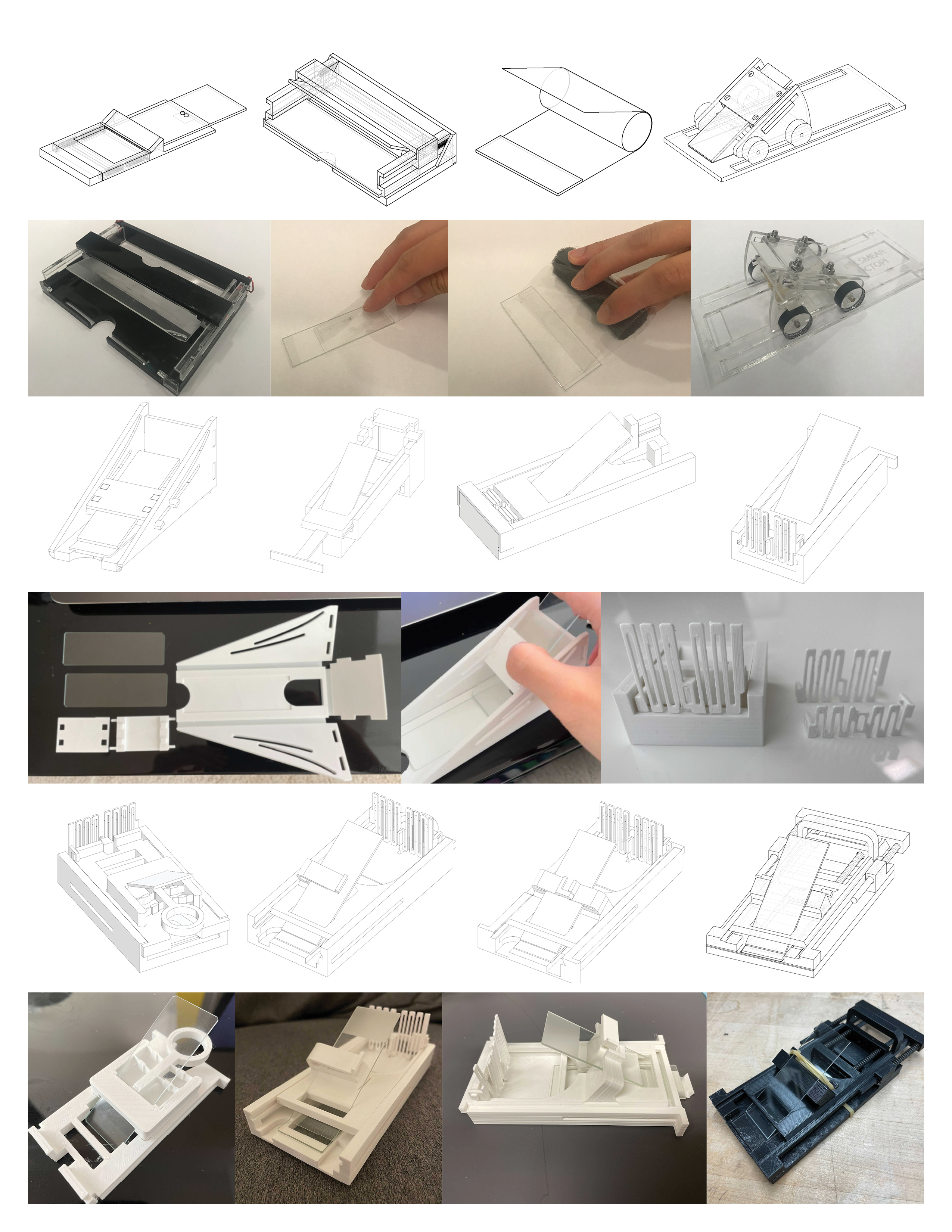}
\caption{\textbf{Early iterations of Inkwell prototype development}
Drawings and photos of early iterations of Inkwell. Explorations included material selection, parameter tolerances, choice of actuator, and fabrication approach. }
\label{figS2}
\end{figure*}

\clearpage

\begin{figure*}[!h]%
\centering
\includegraphics[width=500pt]{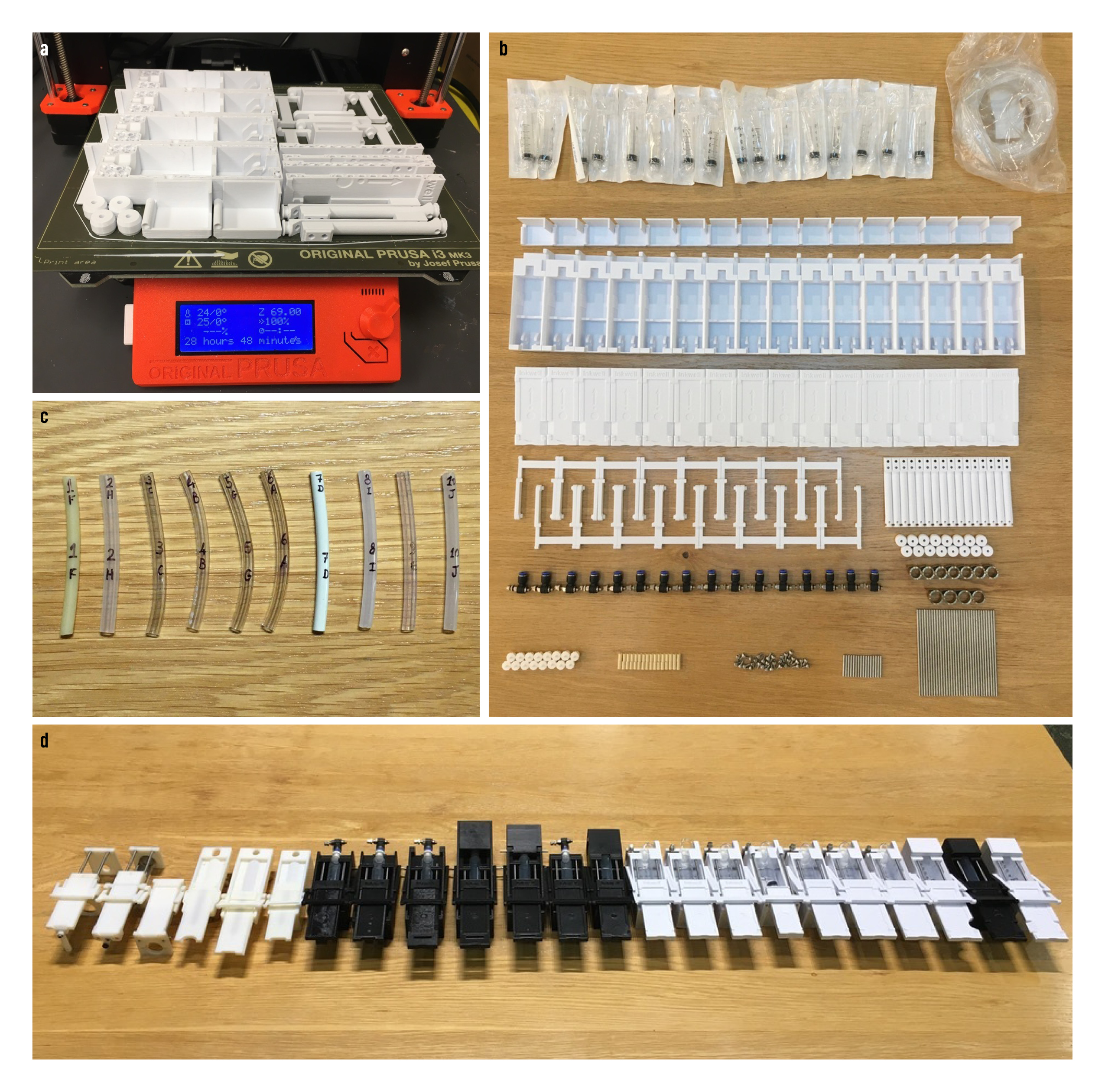}
\caption{\textbf{Late iterations and process images of Inkwell prototype development}
[a] 3D-printing of base components of Inkwell using our in-house filament printer. [b] Raw components for 16 Inkwell units which include 3D-printed elements and off-the-shelf parts. [c] Various tubing material tested for use in Inkwell. [D] Photo of the latest built iterations of Inkwell.}
\label{figS3}
\end{figure*}

\clearpage

\begin{figure*}[!h]%
\centering
\includegraphics[width=400pt]{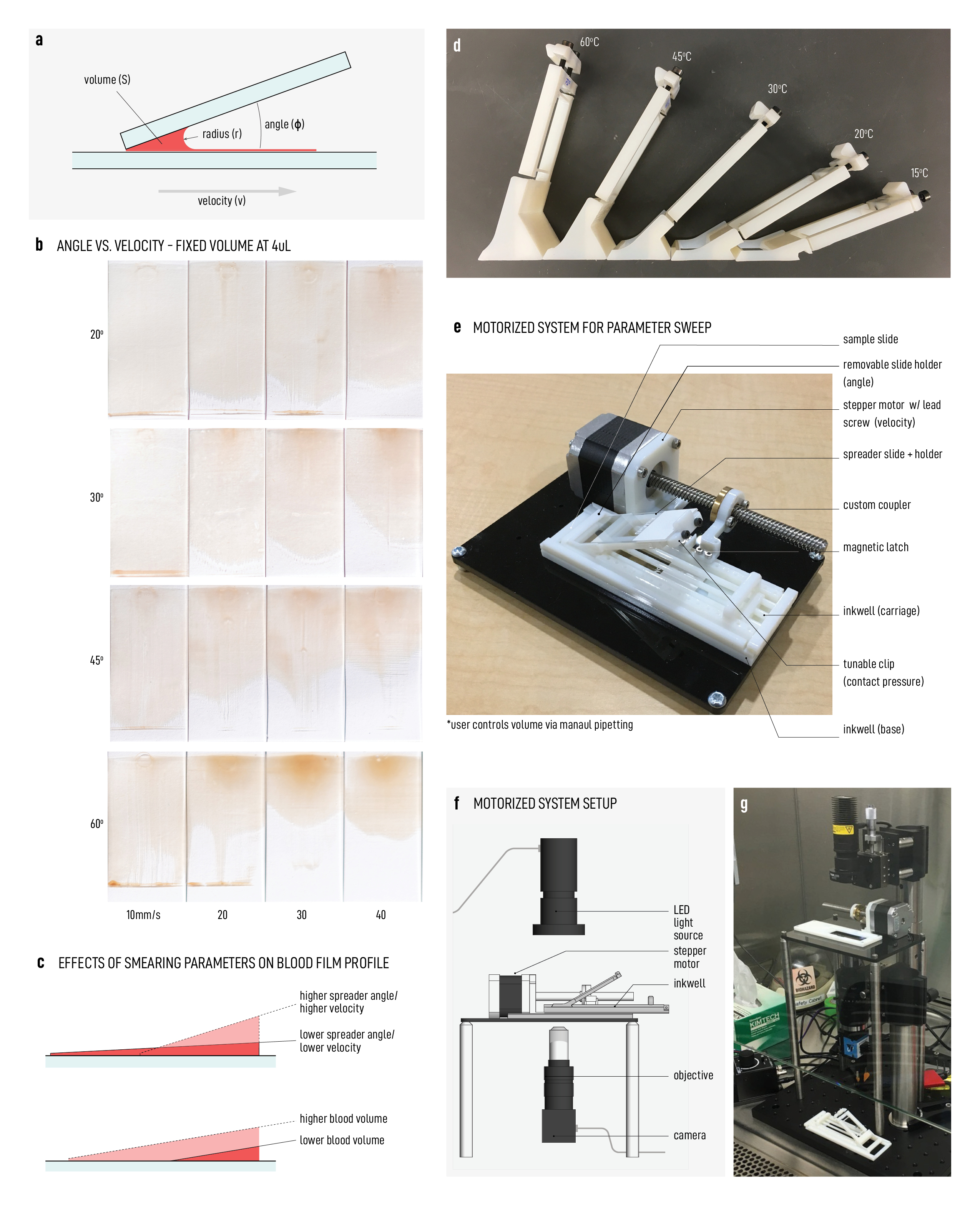}
\caption{\textbf{Experimental setup for identifying ideal parameter values for making good thin smears. }
[a] Schematic diagram showing the various parameters to consider: the initial blood droplet volume, spreader angle, and spreader velocity. [b] An array of thin smears produced using the characterization setup, given a fixed initial blood droplet volume of 4µL while varying the spreader angle(20, 30, 45, 60 degrees) and spreader velocity (10-40mm/s). [c] Schematic sectional diagram depicting the effects of spreader angle, spreader velocity, and initial blood droplet volume on the overall smear profile. We noted that increasing either the spreader angle or velocity yields in a steeper angle in smear profile (meaning blood is deposited at a faster rate relative to the length of the slide. Additionally, decreasing the initial droplet volume yields in a shorter smear length. [d] Interchangeable holder for the spreader slide at various fixed angles (60, 45, 30, 20, 15 degrees) that can be applied onto the motorized characterization setup. [e] Photo of the motorized characterization unit showing the various components that allow for parameters to be controlled. [f] A schematic diagram of the characterization imaging setup. [g] Photo of the characterization imaging setup.}
\label{figS4}
\end{figure*}

\clearpage

\begin{figure*}[!h]%
\centering
\includegraphics[width=500pt]{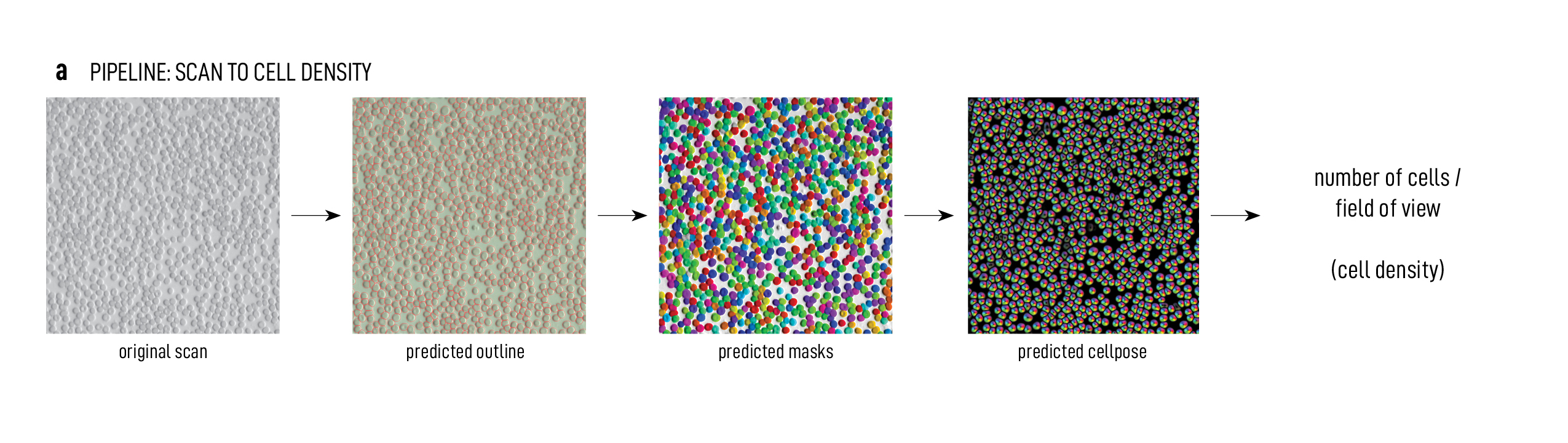}
\caption{\textbf{Pipeline to determine the estimated cell densities}
The Cellpose library was used to estimate the number of cells per field of view from the original scanned image. This method allows us to evaluate the thin smear across the whole slide (as seen in Fig. \ref{fig6}).}
\label{figS5}
\end{figure*}

\clearpage

\begin{figure*}[!h]%
\centering
\includegraphics[width=500pt]{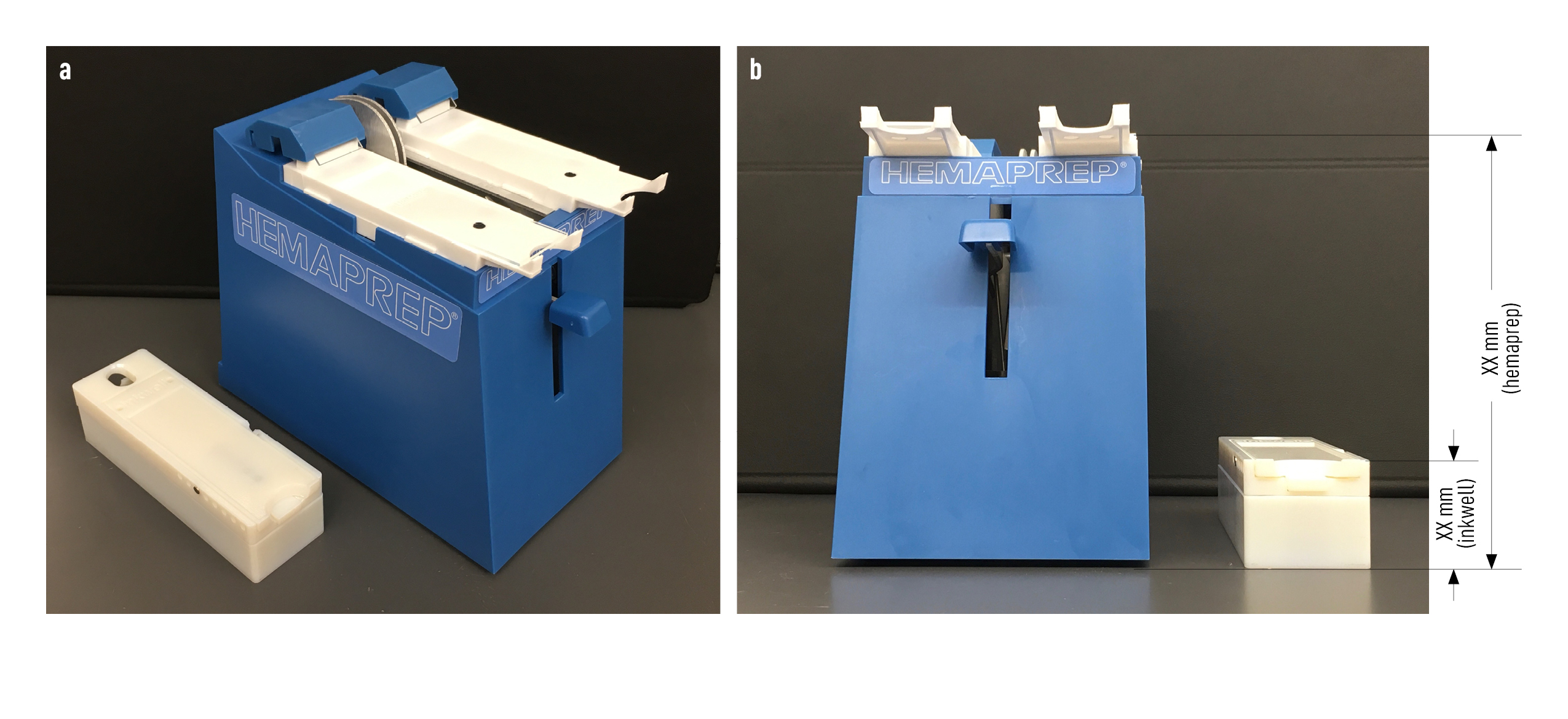}
\caption{\textbf{Scale comparison between Inkwell and Hemaprep}
[a, b] Photos and dimensions showing the scale difference between Inkwell and the industry-standard Hemaprep. Inkwell was designed to be compact and easily portable while having similar performance as Hemaprep.}
\label{figS6}
\end{figure*}

\clearpage

\begin{figure*}[!h]%
\centering
\includegraphics[width=500pt]{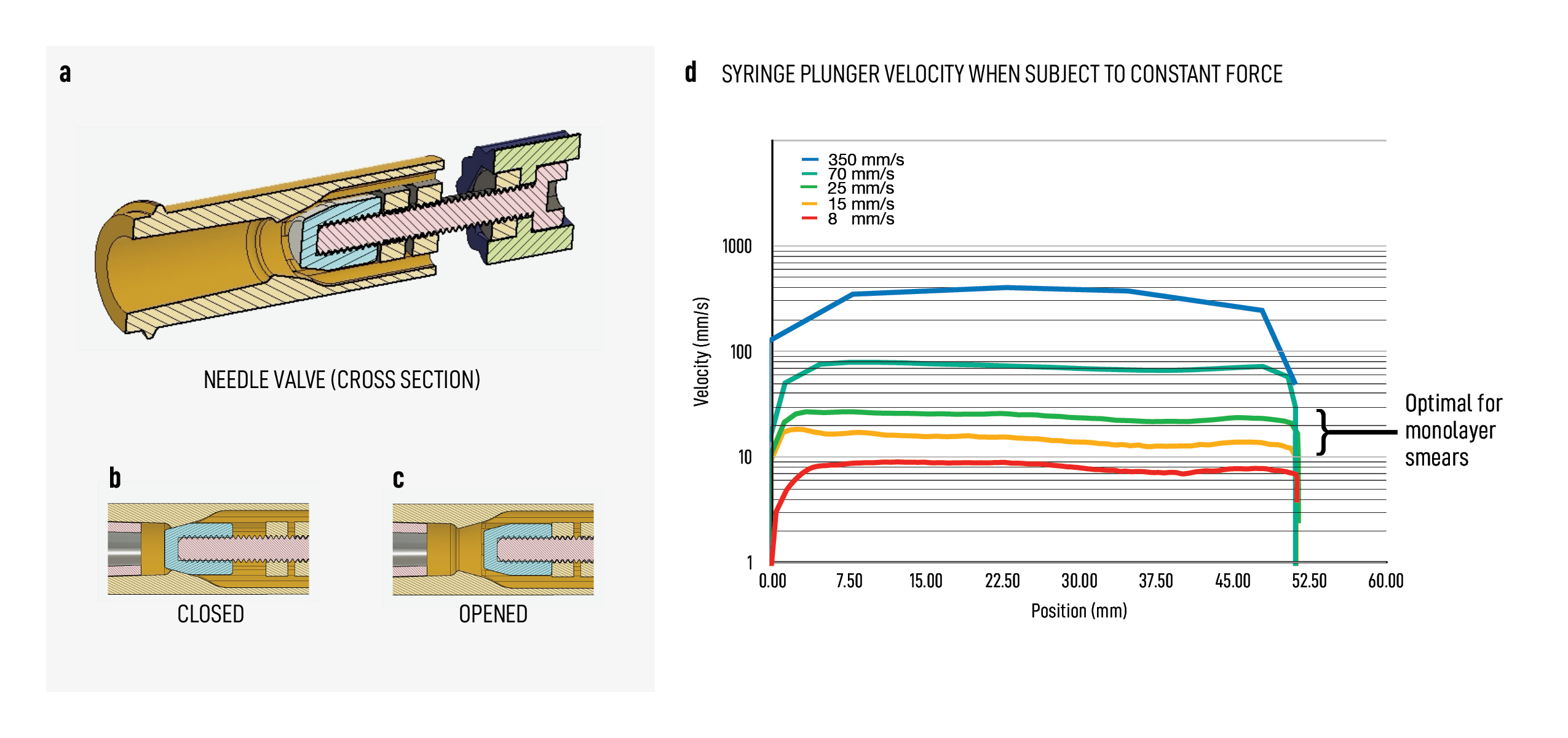}
\caption{\textbf{An alternative custom 3D-printed needle valve to control the velocity of syringe plunger}
[a] Schematic sectional diagrams of an early iteration of a 3D-printed needle valve to be used in Inkwell. It consists of a few 3D-printed parts and standard screws and nuts. In later versions, it was replaced with an equivalent off-the-shelf part. [b] Chart showing the change in syringe plunger velocity over time at various valve tightness settings.}
\label{figS7}
\end{figure*}

\clearpage

\begin{figure*}[!h]%
\centering
\includegraphics[width=500pt]{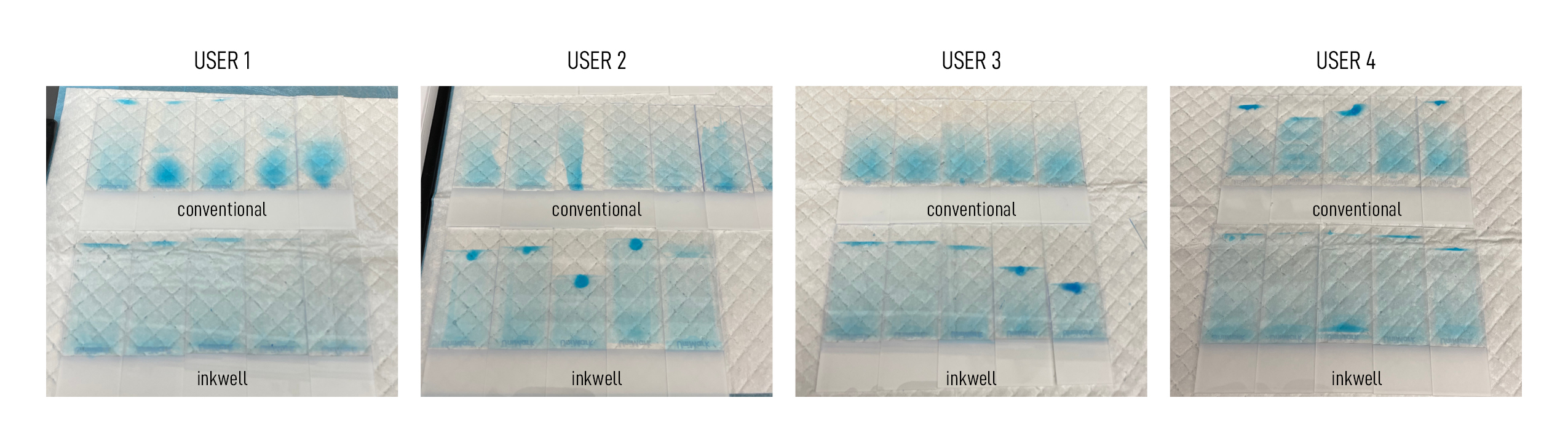}
\caption{\textbf{Smears produced during the Inkwell usability study}
Comparative images of smears produced by each participant done conventionally by hand versus using Inkwell. These smears were done using a phantom blood solution which has similar rheological properties as whole blood.}
\label{figS8}
\end{figure*}

\clearpage

\begin{figure*}[!h]%
\centering
\includegraphics[width=500pt]{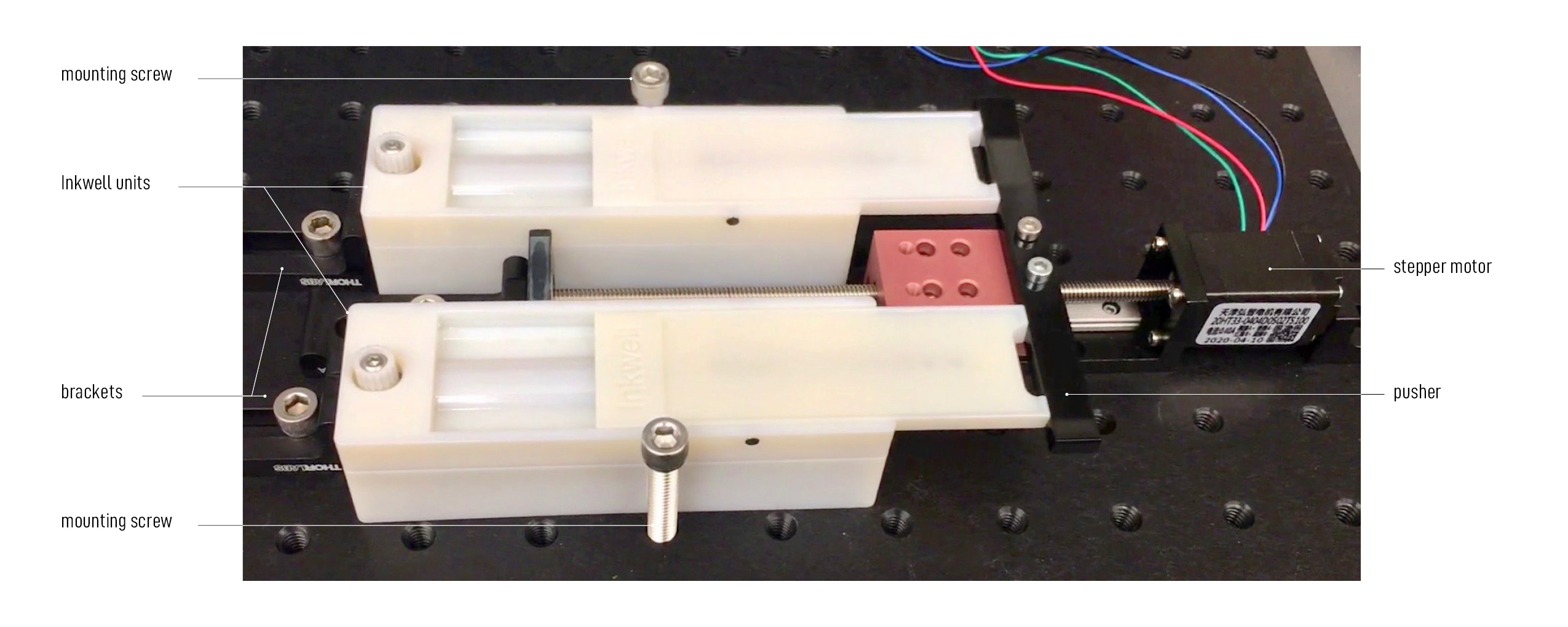}
\caption{\textbf{Inkwell stress test setup}
In order to test the limit of Inkwell by the number of usages, two units were mounted and subjected to numerous push-and-release cycles via a stepper motor to simulate an operator's hand. }
\label{figS9}
\end{figure*}

\begin{figure*}[!h]%
\centering
\includegraphics[width=500pt]{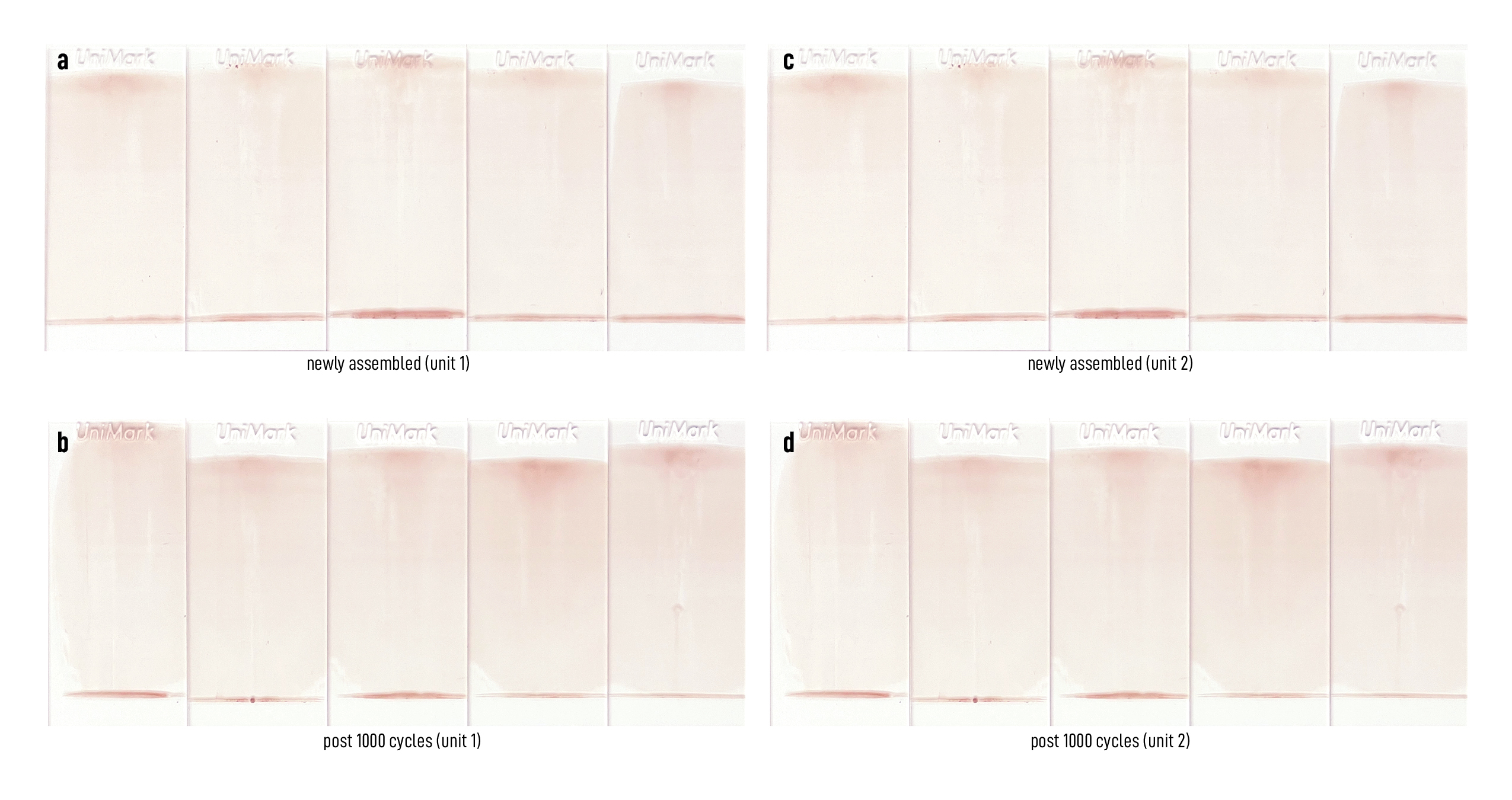}
\caption{\textbf{Additional smears made as part of the robustness testing over 1000 cycles}
Two Inkwell units were used to produce smears using whole blood before and after undergoing 1000 simulated cycles. [a, b] Smears produced by Inkwell unit 1 before and after 1000 cycles. [c, d] Similar smears produced by Inkwell unit 2.}
\label{figS10}
\end{figure*}

\printbibliography

\FloatBarrier
\newpage
\bibliographystyle{naturemag}
\bibliography{main.bib}